\documentclass[twocolumn,superscriptaddress,showpacs,preprintnumbers,amsmath,amssymb,aps,prb]{revtex4}
\usepackage{graphicx}
\usepackage{dcolumn}
\usepackage{bm}
\usepackage{hyperref}

\begin{document}
\title{Effect of significant data loss on identifying electric signals that
 precede rupture by detrended fluctuation analysis in natural time}

\author{E. S. Skordas}
\affiliation{Solid State Section and Solid Earth Physics
Institute, Physics Department, University of Athens,
Panepistimiopolis, Zografos 157 84, Athens, Greece}
\author{N. V. Sarlis}
\affiliation{Solid State Section and Solid Earth Physics
Institute, Physics Department, University of Athens,
Panepistimiopolis, Zografos 157 84, Athens, Greece}
\author{P. A. Varotsos}
\email{pvaro@otenet.gr} \affiliation{Solid State Section and Solid
Earth Physics Institute, Physics Department, University of Athens,
Panepistimiopolis, Zografos 157 84, Athens, Greece}

\begin{abstract}
Electric field variations that appear before rupture have been
recently studied by employing the detrended fluctuation analysis
(DFA) as a scaling method to quantify long-range temporal
correlations. These studies revealed that seismic electric signals
(SES) activities exhibit a scale invariant feature with an
exponent $\alpha_{DFA} \approx 1$ over all scales investigated
(around five orders of magnitude). Here, we study what happens
upon significant data loss, which is a question of primary
practical importance, and show that the DFA applied to the natural
time representation of the remaining data still reveals for SES
activities an exponent close to 1.0, which markedly exceeds the
exponent found in artificial (man-made) noises. This, in
combination with natural time analysis, enables the identification
of a SES activity with probability 75\% even after a significant
(70\%) data loss. The probability increases to 90\% or larger for
50\% data loss.

\end{abstract}

\maketitle

{\bf Complex systems usually exhibit scale-invariant features
characterized by long-range power-law correlations, which are
often difficult to quantify due to various types of
non-stationarities observed in the signals emitted. This also
happens when monitoring geoelectric field changes aiming at
detecting Seismic Electric Signals (SES) activities that appear
before major earthquakes. To overcome this difficulty the novel
method of detrended fluctuation analysis (DFA) has been employed,
which when combined with a newly introduced time domain termed
natural time, allows a reliable distinction of true SES activities
from artificial (man-made) noises. This is so, because the SES
activities are characterized by infinitely ranged temporal
correlations (thus resulting in  DFA exponents close to unity)
while the artificial noises are not. The analysis of SES
observations often meet the difficulty of significant data loss
caused either by failure of the data collection system or by
removal of seriously noise-contaminated data segments. Thus, here
we focus on the study of the effect of significant data loss on
the long-range correlated SES activities quantified by DFA. We
find that the remaining data, even after a considerable percentage
of data loss (which may reach $\sim 80 \%$), may be correctly
interpreted, thus revealing the scaling properties of SES
activities. This is achieved, by applying DFA {\em not} to the
original time series of the remaining data but to those resulted
when employing natural time.}

\section{Introduction}
The output signals of complex systems exhibit fluctuations over
multiple scales\cite{BAS94,MAL95} which are characterized by
absence of dynamic scale, i.e., scale-invariant
behavior\cite{STA95}. These signals, due to the nonlinear
mechanisms controlling the underlying interactions, are also
typically non-stationary and their reliable analysis cannot be
achieved by traditional methods, e.g., power-spectrum and
auto-correlation analysis\cite{HUR51,MAN69,STR81}. On the other
hand, the detrended fluctuation analysis (DFA)\cite{PEN94,TAQ95}
has been established as a robust method suitable for detecting
long-range power-law correlations embedded in non-stationary
signals. This is so, because a power spectrum calculation assumes
that the signal is stationary and hence when applied to
non-stationary time series it can lead to misleading results.
Thus, a power spectrum analysis should be necessarily preceded by
a test for the stationarity of the (portions of the) data
analyzed. As for the DFA, it can determine the (mono) fractal
scaling properties (see below) even in non-stationary time series,
and can avoid, in principle, spurious detection of correlations
that are artifacts of non-stationarities. DFA has been applied
with successful results to diverse fields where scale-invariant
behavior emerges, such as
DNA\cite{PEN93,MAN94,HAV95A,PEN95,HAV95B,MAN96,BUL98,STA99}, heart
dynamics\cite{PEN95B,HO97,IVA99,ASH00,ASH01,IVA01,KAN02,KAR02,IVA04,SCH07,SCH09},
circadian rhythms\cite{HU04,IVA07A,IVA07B,HU07},
meteorology\cite{IVAN99} and climate temperature
fluctuations\cite{BUN98,TAL00,BUN01,MON03,BUN05},
economics\cite{LIU97,VAN97,VAN98,AUS99,VAN99,AUS00,AUS01} as well
as in the low-frequency ($\leq 1$ Hz) variations of the electric
field of the earth that precede
earthquakes\cite{NAT03A,NAT03B,NAT09} termed Seismic Electric
Signals\cite{VAR84A,VAR84B,VAR86,VAR88,VAR91,VAR93A,VAR96A} and in
the relevant\cite{VAR01A,VAR01B,SAR02,PRL03} magnetic field
variations\cite{NAT09}.

Monofractal signals are homogeneous in the sense that they have
the same scaling properties, characterized locally by a single
singularity exponent $h_0$, throughout the signal. Thus,
monofractal signals can be indexed by a single global exponent,
e.g., the Hurst exponent $H \equiv h_0$, which suggests that they
are stationary from viewpoint of their local scaling properties
(see Ref.\cite{IVA01} and references therein). Since DFA can
measure only one exponent, this method is more suitable for the
investigation of monofractal signals. In several cases, however,
the records cannot be accounted for by a single scaling exponent
(i.e., do not exhibit a simple monofractal behavior). In some
examples, there exist crossover (time-) scales separating regimes
with different scaling exponents. In general, if a multitude of
scaling exponents is required for a full description of the
scaling behavior, a multifractal analysis must be applied.
Multifractal signals are intrinsically more complex, and
inhomogeneous, than monofractals (see Ref.\cite{IVA01} and
references therein). A reliable multifractal analysis can be
performed by the Multifractal Detrended Fluctuation Analysis,
MF-DFA\cite{WEB01,KAN02B} or by the wavelet transform (e.g., see
Ref.\cite{MUZ94}).

DFA has been applied, as mentioned, to the SES activities. It was
found\cite{NAT03A} that when DFA is applied to the original time
series of the SES activities and artificial (man-made) noises,
both types of signals lead to a slope at short times (i.e.,
$\Delta t \leq 30s$) lying in the range $\alpha$=1.1-1.4, while
for longer times the range $\alpha$=0.8-1.0 was determined
without, however, any safe classification between SES activities
and artificial noises. On the other hand, when employing natural
time (see Section II), DFA enables the distinction between SES
activities and artificial noises in view of the following
difference: for the SES activities the $\alpha$-values lie
approximately in the range 0.9 - 1.0 (or between 0.85 to 1.1, if a
reasonable experimental error is envisaged), while for the
artificial noises the $\alpha$-values are markedly smaller, i.e.,
$\alpha$=0.65-0.8. In addition, MF-DFA has been
used\cite{NAT03A,NAT03B} and it was found that this multifractal
analysis, when carried out in the conventional time frame, did not
lead to any distinction between these two types of signals, but
does so, if the analysis is made in the natural time domain.

\begin{figure*}
\includegraphics{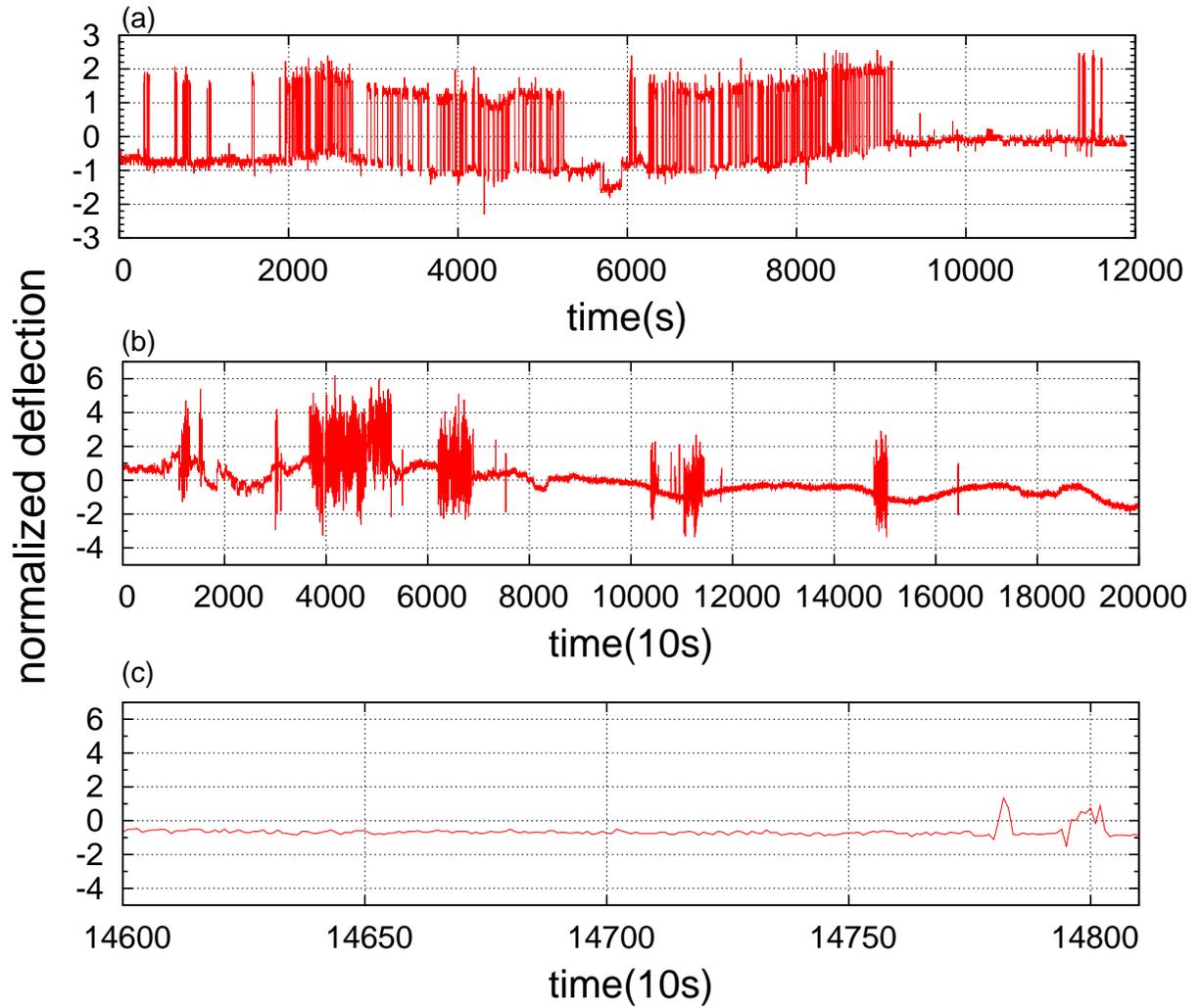}
\caption{(color online)Examples of the electric field recordings
in normalized units, i.e., by subtracting the mean value $\mu$ and
dividing by the standard deviation $\sigma$. The following SES
activities are depicted: (a)the one recorded on April 18, 1995 at
Ioannina station, (b)the long duration SES activity recorded from
December 27, 2010 to  December 30, 2009 at Lamia station. (c) is
an excerpt of (b) showing that, after long periods of quiescence,
the electric field exhibits measurable excursions (transient
pulses).} \label{fig1}
\end{figure*}

\begin{figure*}
\includegraphics{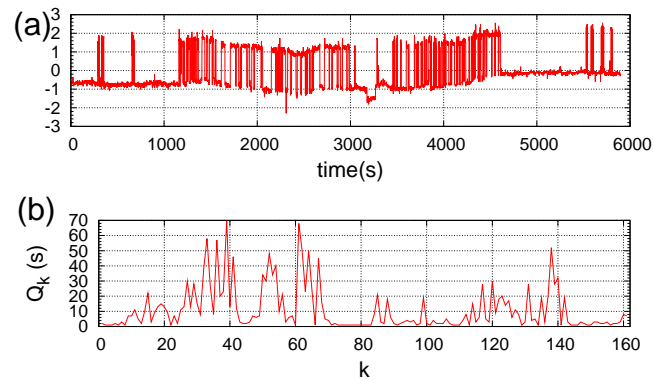}
\caption{(color online)(a):Example of a surrogate time-series (in
normalized units as in Fig.\ref{fig1}) obtained by removing
segments of length $L=200$ from the signal of Fig.\ref{fig1}(a)
with 50\% data loss (i.e., $p=0.50$). (b):The natural time
representation of (a). The values obtained from the analysis of
(b) in natural time are $\kappa_1=0.067(4)$, $S=0.076(4)$,
$S_-=0.071(4)$ and $a_{DFA}=0.90(5)$.}\label{figx}
\end{figure*}

\begin{figure}
\includegraphics{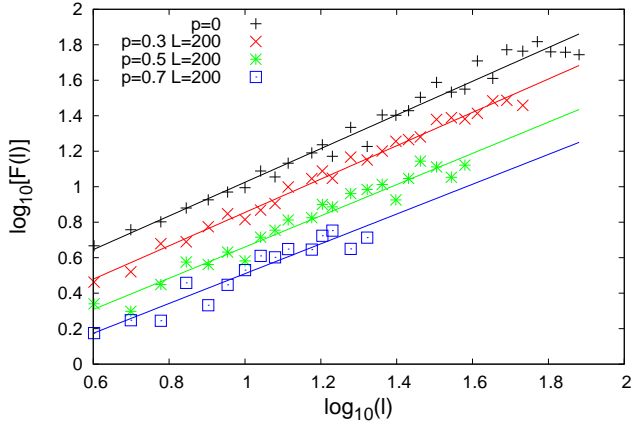}
\caption{(color online) The dependence of the DFA measure $F(l)$
versus the scale $l$ in natural time: we increase the percentage
of data loss p by removing segments of length $L=200$ samples from
the signal of Fig.\ref{fig1}(a). The black (plus) symbols
correspond to no data loss (p=0), the red (crosses) to 30\% data
loss (p=0.3), the green (asterisks) to 50\% data loss (p=0.5) and
the blue (squares) to 70\% data loss (p=0.7).   Except for the
case p=0, the data  have been shifted vertically for the sake of
clarity. The slopes of the corresponding straight lines that fit
the data lead to $\alpha_{DFA}=$0.95, 0.94, 0.88 and 0.84 from the
top to bottom, respectively. They correspond to the average values
of $\alpha_{DFA}$ obtained from 5000 surrogate time-series that
were generated with the method of surrogate by Ma et
al.\cite{MA10} (see the text). } \label{fig2}
\end{figure}

\begin{figure}
\includegraphics{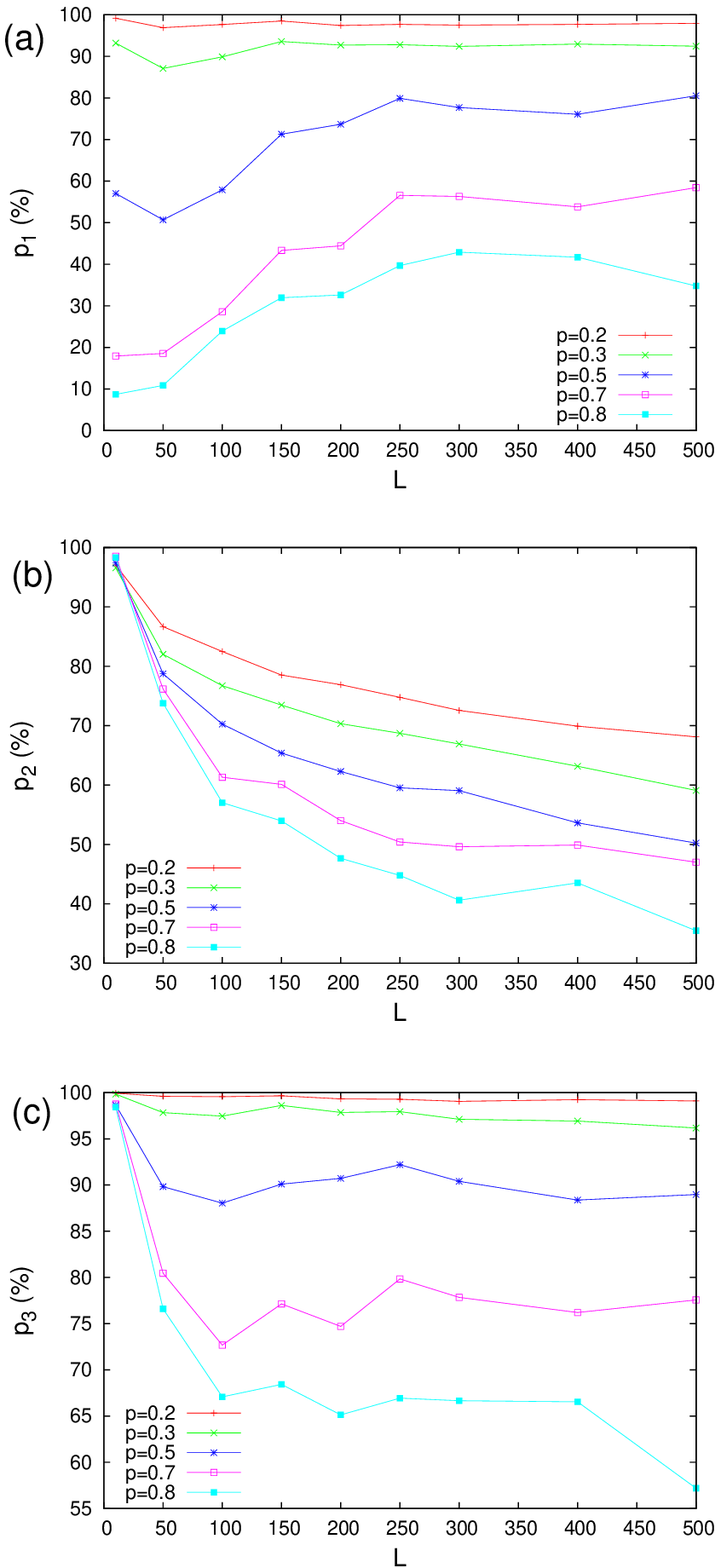}
\caption{(color online) The probabilities $p_1$(a), $p_2$(b) and
$p_3$(c)  to recognize the signal of Fig.\ref{fig1}(a) as true SES
activity when considering various percentages of data loss p=0.2,
0.3, 0.5, 0.7 and 0.8 as a function of the length $L$ of the
contiguous samples removed. The removal of large segments leads to
better results when using DFA in natural time (a), whereas the
opposite holds when using the conditions of Eqs.(\ref{eq1}) and
(\ref{eq2}) for  $\kappa_1$, $S$ and $S_-$ (b). The optimum
selection (c) for the identification of a signal as SES activity
consists of a proper combination of the aforementioned procedures
in (a) and (b), see the text. The values presented have been
obtained from 5000 surrogate time-series (for a given value of $p$
and $L$), and hence they a have plausible error 1.4\% ($\approx
1/\sqrt{5000}$). } \label{fig3}
\end{figure}

The aforementioned findings of DFA for SES activities, are
consistent with their generation mechanism\cite{VARBOOK,NEWBOOK}
which could be summarized as follows. Beyond the usual intrinsic
lattice defects\cite{VAR76,VAR78A,VAR78B,VAR79A,VAR84C,VAR07} that
exist in solids, in ionic solids in particular, when doped with
aliovalent impurities, extrinsic defects are formed for the sake
of charge compensation. A portion of these defects are attracted
by the nearby impurities, thus forming electric dipoles the
orientation of which can change by means of a defect migration. In
the focal area of an impending earthquake the stress gradually
increases and hence affects the thermodynamic parameters of this
migration, thus it may result in a gradual decrease of their
relaxation time. When the stress (pressure) reaches a {\em
critical} value, a {\em cooperative} orientation of these dipoles
occurs, which leads to the emission of a transient signal. This
signal constitutes the SES activity and, since it is characterized
by {\em critical} dynamics, should exhibit infinitely range
temporal correlations. This is consistent with the above findings
of DFA that for SES activities $\alpha_{DFA} \approx 1$.

Hereafter, we will solely use DFA in view of its simplicity and
its ability to reliably classify SES activities. It is the basic
aim of this study to investigate how significant data loss affects
the scaling behavior of long-range correlated SES activities
inspired from the new segmentation approach introduced recently by
Ma et al\cite{MA10} to generate surrogate signals by randomly
removing data segments from stationary signals with different
types of long-range correlations.  The practical importance of
this study becomes very clear upon considering that such a data
loss is inevitable mainly due to the following two reasons: First,
failure of the measuring system in the field station, including
the electric measuring dipoles, electronics and the data
collection system, may occur especially due to lightning. Second,
noise-contaminated data segments are often unavoidable due to
natural changes such as rainfall, lightning, induction of
geomagnetic field variations and ocean-earth tides besides the
noise from artificial (man-made) sources including the leakage
currents from DC driven trains. The latter are common in Japan
where at some sites they may last for almost 70$\%$ of the time
every day. We clarify, however, that even at such noisy-stations
in Japan, several clear SES activities have been unambiguously
identified\cite{ORI09} during the night (when the noise level is
low). In addition, prominent SES activities were recently
reported\cite{UYE09} at noise-free stations (far from
industrialized regions) having long duration, i.e., of the order
of several weeks. As we shall see, our results described in
Section IV, are in essential agreement with those obtained in the
innovative and exhaustive study of Ma et al\cite{MA10}. Before
proceeding to our results, we will briefly summarize DFA and
natural time analysis in Section II, and then present in Section
III the most recent SES data along with their analysis in natural
time. In Section V, we summarize our conclusions.

\section{Conventional detrended fluctuation analysis. Natural time.}
We first sum up the original time series and determine the profile
$y(i), i=1,\ldots,N$; We then divide this profile of length $N$
into $N/l$($ \equiv N_l$) non overlapping fragments of
$l$-observations. Next, we define the detrended process $y_{l,\nu}
(m)$, in the  $\nu$-th fragment, as the difference between the
original value of the profile and the local (linear) trend. We
then calculate the mean variance of the detrended process:
\begin{equation}\label{eqdfa1}
F^2(l)=\frac{1}{N_l}\sum_{\nu=1}^{N_l} f^2(l,\nu)
\end{equation}
where
\begin{equation}\label{eqdfa2}
f^2(l,\nu)=\frac{1}{l} \sum^l_{m=1} y^2_{l,\nu}(m)
\end{equation}
If $F(l) \sim l^\alpha$, the slope of the log$F(l)$ versus log$l$
plot, leads to the value of the exponent $\alpha_{DFA}\equiv
\alpha$. (This scaling exponent is a self-similarity parameter
that represents the long-range power-law correlations of the
signal). If $\alpha_{DFA}$=0.5, there is no correlation and the
signal is uncorrelated (white noise); if $\alpha_{DFA} <$0.5, the
signal is anti-correlated; if  $\alpha_{DFA} >$0.5, the signal is
correlated and specifically the case  $\alpha_{DFA}$=1.5
corresponds to the Brownian motion (integrated white noise).

We now summarize the background of natural time $\chi$. In a time
series comprising $N$ events, the natural time $\chi_k = k/N$
serves as an index\cite{NAT02} for the occurrence of the $k$-th
event. The evolution of the pair ($\chi_k, Q_k$) is
studied\cite{NAT02,NAT03A,NAT03B,NAT04,NAT05B,NAT05A,NAT06A,NAT06B},
where $Q_k$ denotes a quantity proportional to the {\em energy}
released in the $k$-th event.
 For dichotomous
signals, which is frequently the case of SES activities, the
quantity $Q_k$ stands for the duration of the $k$-th pulse. By
defining $p_k=Q_{k}/\sum_{n=1}^{N}Q_{n}$,  we have
found\cite{NAT02} that the variance
\begin{equation}
\kappa_1=\langle \chi^2 \rangle -\langle \chi \rangle ^2,
\end{equation}
 where $\langle f( \chi) \rangle = \sum_{k=1}^N p_k f(\chi_k )$,
 of the natural time $\chi$ with respect to the distribution $p_k$
 may be used\cite{NAT03A,NAT03B} for the identification of SES
 activities.  In particular, the following relation should hold
\begin{equation} \label{eqk1}
 \kappa_1\approx 0.070.
 \end{equation}

  The entropy $S$ in the natural time-domain is defined
as\cite{NAT03B} \begin{equation}
 S \equiv  \langle \chi \ln \chi
\rangle - \langle \chi \rangle \ln \langle \chi
\rangle.\end{equation} It exhibits\cite{NAT05B}
Lesche\cite{LES82,LES04} (experimental) stability, and for SES
activities (critical dynamics) is smaller\cite{NAT03B} than the
value $S_u (=\ln 2 /2-1/4\approx 0.0966$) of a ``uniform'' (u)
distribution (as defined in Refs. \cite{NAT03A,NAT03B,NAT04}, e.g.
when all $p_k$ are equal or $Q_k$ are positive independent and
identically distributed random variables of finite variance. In
this case, $\kappa_1$ and $S$ are designated $\kappa_u(=1/12)$ and
$S_u$, respectively.). Thus, $S < S_u$. The same holds for the
value of the entropy obtained\cite{NAT05B,NAT06A} upon considering
the time reversal $\hat{T}$, (the operator $\hat{T}$ is defined by
${\hat{T}} p_k=p_{N-k+1}$), which is labelled by $S_-$.

In summary, the SES activities, in contrast to the signals
produced by man-made electrical sources, when analyzed in natural
time (see Section I) exhibit {\em infinitely} ranged temporal
correlations\cite{NAT02,NAT03B} and obey the
conditions\cite{NAT06A}:
\begin{equation}\label{eq1}
    \kappa_1 \approx 0.07
\end{equation}
and
\begin{equation}\label{eq2}
    S, S_- < S_u.
\end{equation}

 It should be recalled that SES activities are publicized
{\em only} when the magnitude of the impending earthquake is
Ms(ATH)$\geq$6.0

\section{The experimental data}
In Fig.\ref{fig1}(a), we depict the SES activity recorded at
Ioannina station (Northwestern Greece) on 18 April, 1995. It
preceded the 6.6 earthquake on 13 May, 1995. Since this earthquake
was the strongest one in Greece during the 25 year period
1983-2007, we focus on this example in the next Section  to
present the DFA results.

In addition, the most recent SES activity in Greece is depicted in
Figs.\ref{fig1}(b),(c). This has been recorded at Lamia station
located in Central Greece during the period 27 December - 30
December, 2009. Almost three weeks later, two strong earthquakes
of magnitude $M_s$(ATH)=5.7 and 5.6 occurred in central Greece
with an epicenter at 38.4$^o$N 22.0$^o$E (but see also
Refs.\cite{ARXIV10,keimeno}).

The two signals in Figs. \ref{fig1}(a),(b) have been classified as
SES activities after analyzing them in natural time. In
particular, for the signal in Fig. \ref{fig1}(a), straightforward
application of natural time analysis leads to the conclusion that
the conditions (\ref{eq1}) and (\ref{eq2}) are satisfied (see
Table I of Ref.\cite{NAT05B}). Further, the DFA analysis of the
natural representation of this signal gives an exponent
$\alpha_{DFA}=0.95(4) \approx 1$. (We also note that the
classification of this signal as SES activity had been previously
achieved by independent procedures discussed\cite{VAR96B} in the
Royal Society Meeting that was held during May 11-12, 1995
before\cite{LIG96} the occurrence of the 6.6 earthquake on May 13,
1995.) For the long duration signal of Fig. \ref{fig1}(b), the
procedure explained in detail in Ref.\cite{NAT09} was followed.

\section{Data analysis and results}
Following Ma et al\cite{MA10}, we now describe the segmentation
approach used here to generate surrogate signals $\tilde{u}(i)$ by
randomly removing data segments of length $L$ from the original
signal $u(i)$. The percentage $p$ of the data loss, i.e., the
percentage of the data removed, also characterizes the signal
$\tilde{u}(i)$. The procedure followed is based on the
construction of a binary time-series $g(i)$ of the same length as
$u(i)$. The values of $u(i)$ that correspond to $g(i)$ equal to
unity are kept, whereas the data of $u(i)$ when $g(i)$ equals zero
are removed. The values of $u(i)$ kept, are then concatenated to
construct $\tilde{u}(i)$.

The binary time-series $g(i)$ is obtained as follows\cite{MA10}:
(i)We first generate the lengths $l_j=L$ with $j=1,2,\ldots ,M$ of
the removed segments, by selecting $M$ to be the smallest integer
so that the total number of removed data satisfies the condition
$\sum_{j=1}^{M}l_j\geq pN$. (ii) We then construct an auxiliary
time-series $a(k)$ with $a(k)=L$ when $k=1,2,\ldots ,M$ and
$a(k)=1$ when $k=M+1, \ldots, N-M(L+1)$ of size $N-M(L+1)$.
(iii)We shuffle the time-series $a(k)$ randomly to obtain
$\tilde{a}(k)$. (iv)We then append $\tilde{a}(k)$ to obtain
$g(i)$: if $\tilde{a}(k)=1$ we keep it, but we replace all
$\tilde{a}(k)=L$ with $L$ elements  of value '0' and one element
with value '1'. In this way, a binary series $g(i)$ is obtained,
which has a size equal to the one of the original signal $u(i)$.
We then construct the surrogate signal $\tilde{u}(i)$ by
simultaneously scanning the original signal $u(i)$ and the binary
series $g(i)$, removing the $i$-th element of $u(i)$ if $g(i)=0$
and concatenating the segments of the remaining data to
$\tilde{u}(i)$.

The resulting signal $\tilde{u}(i)$ is later analyzed in natural
time, thus leading to the quantities $\kappa_1$, $S$ and $S_-$ as
well as to the DFA exponent $\alpha_{DFA}$ in natural time. Such
an example is given in Fig. \ref{figx}. In what remains we present
the results focusing hereafter, as mentioned, on the example of
the SES activity depicted in Fig. \ref{fig1}(a).

Typical DFA plots, obtained for $L$=200 and $p=30, 50$ and $70 \%$
are given in Fig. \ref{fig2}. For the sake of comparison, this
figure also includes the case of no data loss (i.e., $p = 0$). We
notice a gradual decrease of $\alpha_{DFA}$ upon increasing the
data loss, which affects our ability to classify a signal as SES
activity.

In order to quantify, in general, our ability to identify SES
activities from the natural time analysis of surrogate signals
with various levels of data loss, three procedures have been
attempted:

 Let us call procedure 1, the investigation
whether $\alpha_{DFA}$, resulted from the DFA analysis  of the
natural time representation of a signal, belongs to the range
$0.85 \leq \alpha_{DFA} \leq 1.0$. If it does, the signal is then
classified as SES activity. Figure \ref{fig3}(a) shows that for a
given amount of data loss ($p$=const), upon increasing the length
$L$ of the randomly removed segment, the probability $p_1$ of
achieving, after making 5000 attempts (for a given value of $p$
and $L$), the identification of the signal as SES activity is
found to gradually increase versus $L$ at small scales and
stabilizes at large scales. For example, when considering the case
of $70 \%$ data loss (magenta color in Fig. \ref{fig3}(a)) the
probability $p_1$ is close to $20 \%$ for $L$=50; it increases to
$p_1 \approx$ $30 \%$ for $L$=100 and finally stabilizes around
$50 \%$ for lengths $L=$300 to 500.

Let us now label as procedure 2, the investigation whether the
quantities $\kappa_1$, $S$ and $S_-$ (resulted from the analysis
of a signal in natural time) obey the conditions (\ref{eq1}) and
(\ref{eq2}), i.e., $|\kappa_1-0.070|\leq 0.01$ and $S$, $S_-$
$<$0.0966. If they do so, the signal is classified as SES
activity. Figure \ref{fig3}(b) shows that for a given amount of
data loss, the probability $p_2$ of achieving the signal
identification as SES activity -that results after making 5000
attempts for each $L$ value- gradually decreases when moving from
the small to large scales. Note that for the smallest length scale
investigated, i.e., $L$=10 (which is more or less comparable -if
we consider the sampling frequency of 1 sample/sec- with the
average duration $\approx$11 sec of the transient pulses that
constitute the signal), the probability $p_2$ reaches values close
to $100 \%$ even for the extreme data loss of $80 \%$. This is
understood in the context that the quantities $\kappa_1$, $S$ and
$S_-$ remain almost unaffected when randomly removing segments
with lengths comparable to the average pulse's duration. This is
consistent with our earlier finding\cite{NAT05B} that the
quantities $\kappa_1$, $S$ and $S_-$ are experimentally stable
(Lesche's stability) meaning that they exhibit only slight
variations when deleting (due to experimental errors) a small
number of pulses. On the other hand, at large scales of $L$, $p_2$
markedly decreases. This may be understood if we consider that, at
such scales, each segment of contiguous $L$ samples removed,
comprises on the average a considerable number of pulses the
removal of which may seriously affect the quantities $\kappa_1$,
$S$ and $S_-$. As an example, for $80 \%$ data loss (cyan curve in
Fig. \ref{fig3}(b)), and for lengths $L$=400-500, the probability
$p_2$ of identifying a true SES activity is around $40 \%$.

Interestingly, a closer inspection of the two figures
\ref{fig3}(a) and \ref{fig3}(b) reveals that $p_1$ and $p_2$ play
complementary roles. In particular, at small scales of $L$, $p_1$
increases but $p_2$ decreases versus $L$. At large scales, where
$p_1$ reaches (for considerable values of data loss) its largest
value, the $p_2$ value becomes small. Inspired from this
complementary behavior of $p_1$ and $p_2$, we proceeded to the
investigation of a combined procedure, let us call it procedure 3.
In this procedure, a signal is identified as SES activity when
{\em either} the condition $0.85 \leq \alpha_{DFA} \leq 1.0$ {\em
or} the relations (\ref{eq1}) and (\ref{eq2}) are satisfied. The
probability $p_3$ of achieving such an identification, after
making 5000 attempts (for a given value of $p$ and $L$), is
plotted in Fig. \ref{fig3}(c). The results are remarkable since,
even at significant values of data loss, e.g., $p= 70 \%$ or $80
\%$, the probability $p_3$ of identifying a SES activity at scales
$L$=100 to 400 remains relatively high, i.e., $p_3 \approx 75 \%$
and $65 \%$, respectively (cf. note also that the value of $p_3$
reaches values close to $100 \%$ at small scales $L$=10). This is
important from practical point of view, because it states for
example the following: Even if the records of a station are
contaminated by considerable noise, say $70 \%$ of the time of its
operation, the remaining $30 \%$ of the non-contaminated segments
have a chance of $\sim 75 \%$ to correctly identify a SES
activity. The chances increase considerably, i.e., to $p_3 \approx
90 \%$, if only half of the recordings are noisy.

The aforementioned results have been deduced from the analysis of
a SES activity lasting around three hours. In cases of SES
activities with appreciably longer duration, e.g., a few to
several days\cite{SAR08,NAT09} detected in Greece or a few months
in Japan\cite{UYE09}, the results should become appreciably
better.

\begin{figure*}
\includegraphics{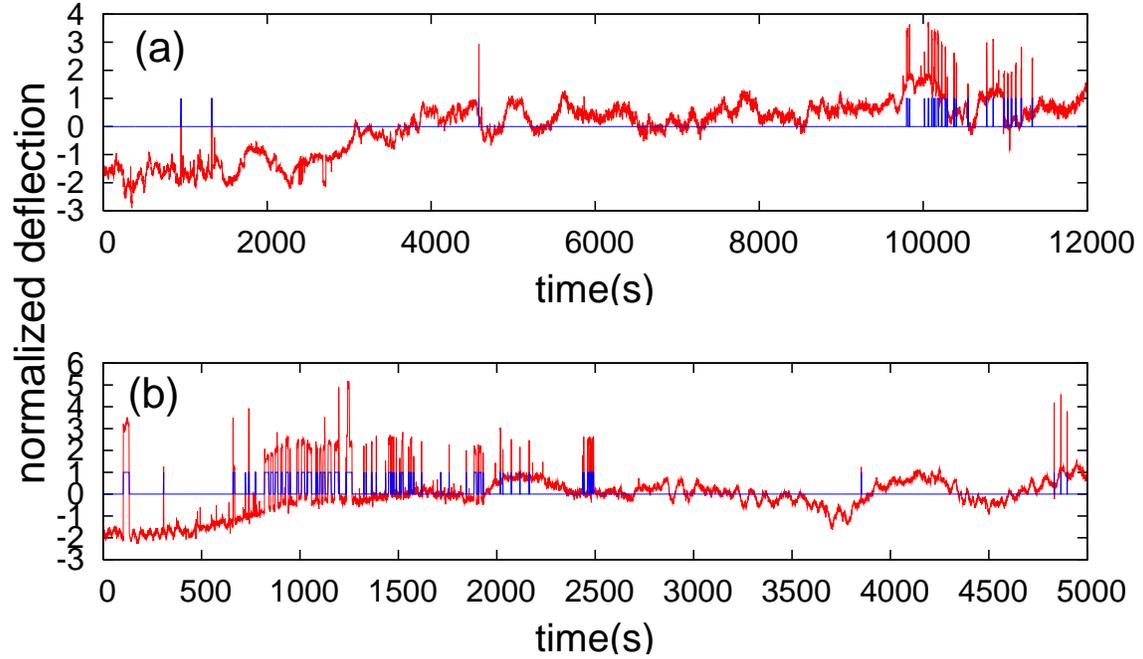}
\caption{(color online)Electric field recordings at PAT in
normalized units, i.e., by subtracting the mean value $\mu$ and
dividing by the standard deviation $\sigma$. The following SES
activities are depicted in red: (a) on August 9, 2010, (b) on
August 10, 2010. The dichotomous representations (blue) lead to
the following values: $\kappa_1=0.072(4)$, $S=0.068(6)$,
$S_-=0.088(5)$ and $\kappa_1=0.064(5)$, $S=0.088(8)$,
$S_-=0.062(3)$, for (a) and (b) respectively. } \label{fig5}
\end{figure*}

\begin{figure*}
\includegraphics{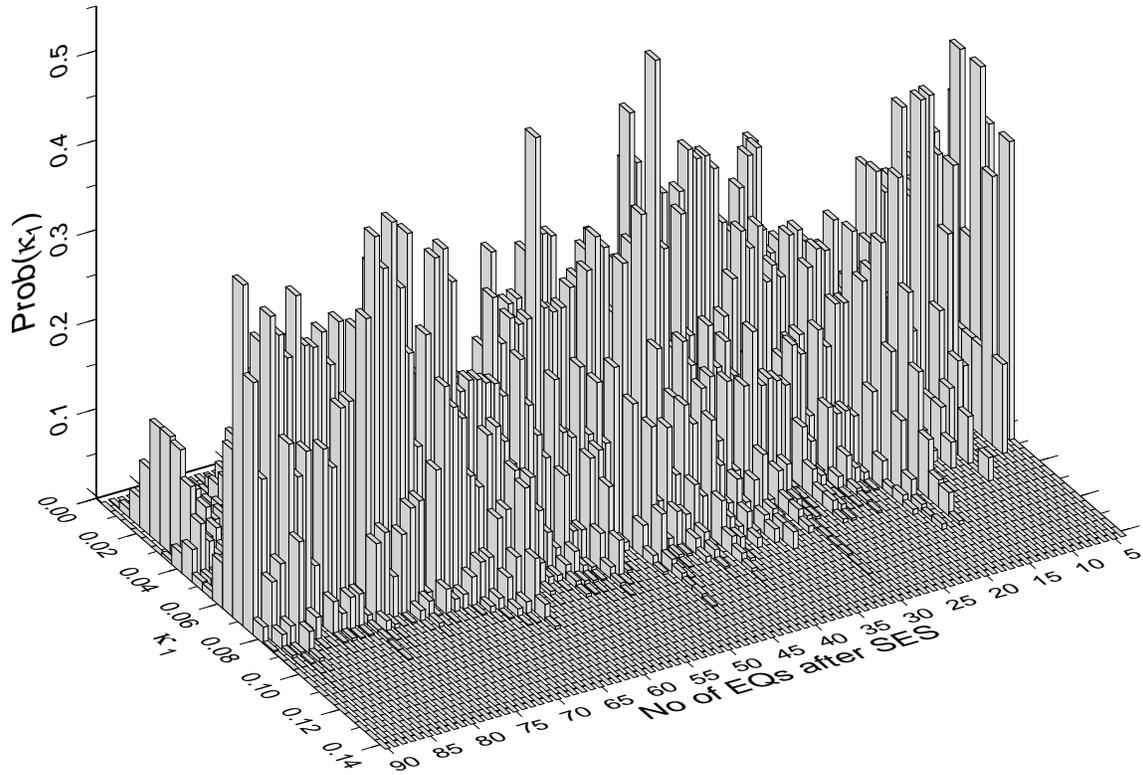}
\caption{The probability Prob($\kappa_1$) vs $\kappa_1$ upon considering
the seismicity (M$_{thres}$=3.1) after August 10, 2010  until 15:42 UT on September 29,
2010 within the area $N_{37.5}^{38.8}E_{19.8}^{23.3}$.} \label{fig6}
\end{figure*}

\section{Conclusions}
We start our conclusions by recalling that the distinction between
SES activities (critical dynamics, infinitely ranged temporal
correlations) and artificial (man-made) noise remains an extremely
difficult task, even without any data loss, when solely focusing
on the original time series of electrical records which are, of
course, in conventional time. On the other hand, when combining
natural time with DFA analysis, such a distinction becomes
possible even after significant data loss. In particular we showed
for example that even when randomly removing $50 \%$ of the data,
we have a probability ($p_3$) around $90 \%$, or larger, to
identify correctly a SES activity. This probability becomes
somewhat smaller, i.e., 75\%, when the data loss increases to
70\%. To achieve this goal, the proper procedure is the following:
the signal is first represented in natural time and then analyzed
in order to deduce the quantities $\kappa_1$, $S$ and $S_-$ as
well as the exponent $\alpha_{DFA}$ from the slope of the log-log
plot of the DFA analysis in natural time. We then examine whether
the latter slope has a value close to unity {\em or} the
conditions $\kappa_1 \approx 0.070$ and $S$, $S_-$ $< S_u$ are
obeyed. In other words, the consequences caused by an undesirable
severe data loss can be markedly reduced upon taking advantage of
the DFA and natural time analysis.

\begin{figure*}
\includegraphics{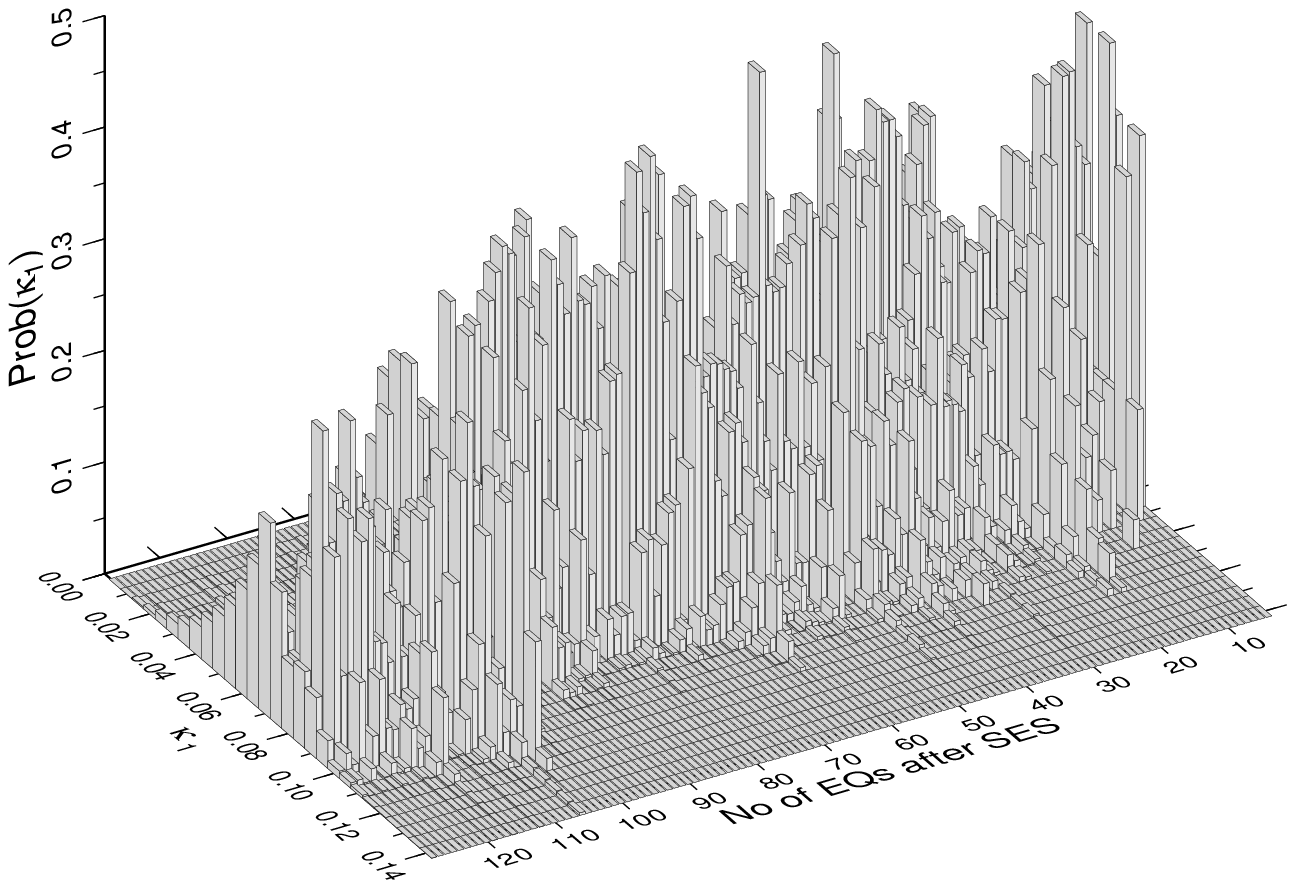}
\caption{The probability Prob($\kappa_1$) vs $\kappa_1$ upon considering
the seismicity (M$_{thres}$=3.1) after August 10, 2010  until 04:04 UT on October 28,
2010 within the area $N_{37.5}^{38.8}E_{19.8}^{23.3}$.} \label{fig7}
\end{figure*}

\begin{figure*}
\includegraphics[scale=0.8]{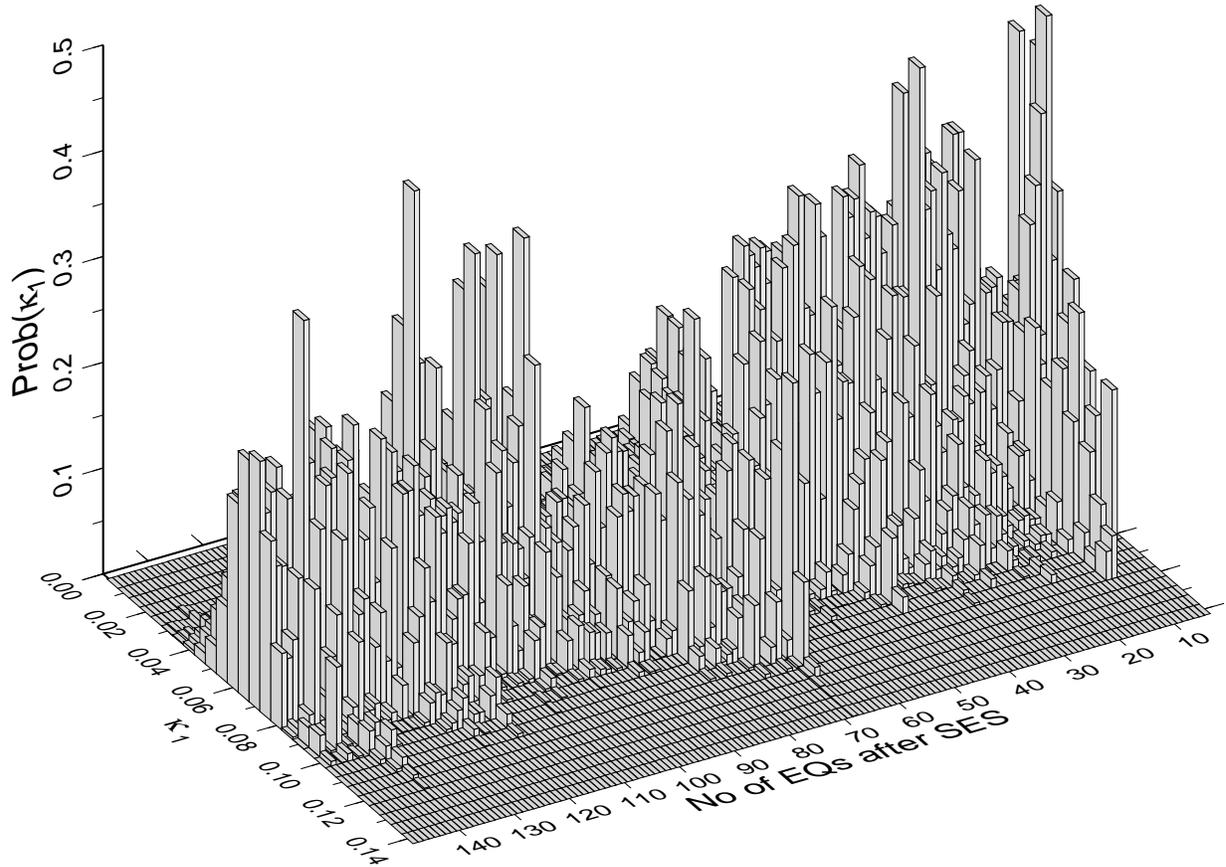}
\caption{Continuation of the study of the analysis of seismicity in natural time as described in the Note added on January 27, 2011.} \label{fig8}
\end{figure*}

\begin{figure*}
\includegraphics[scale=0.8]{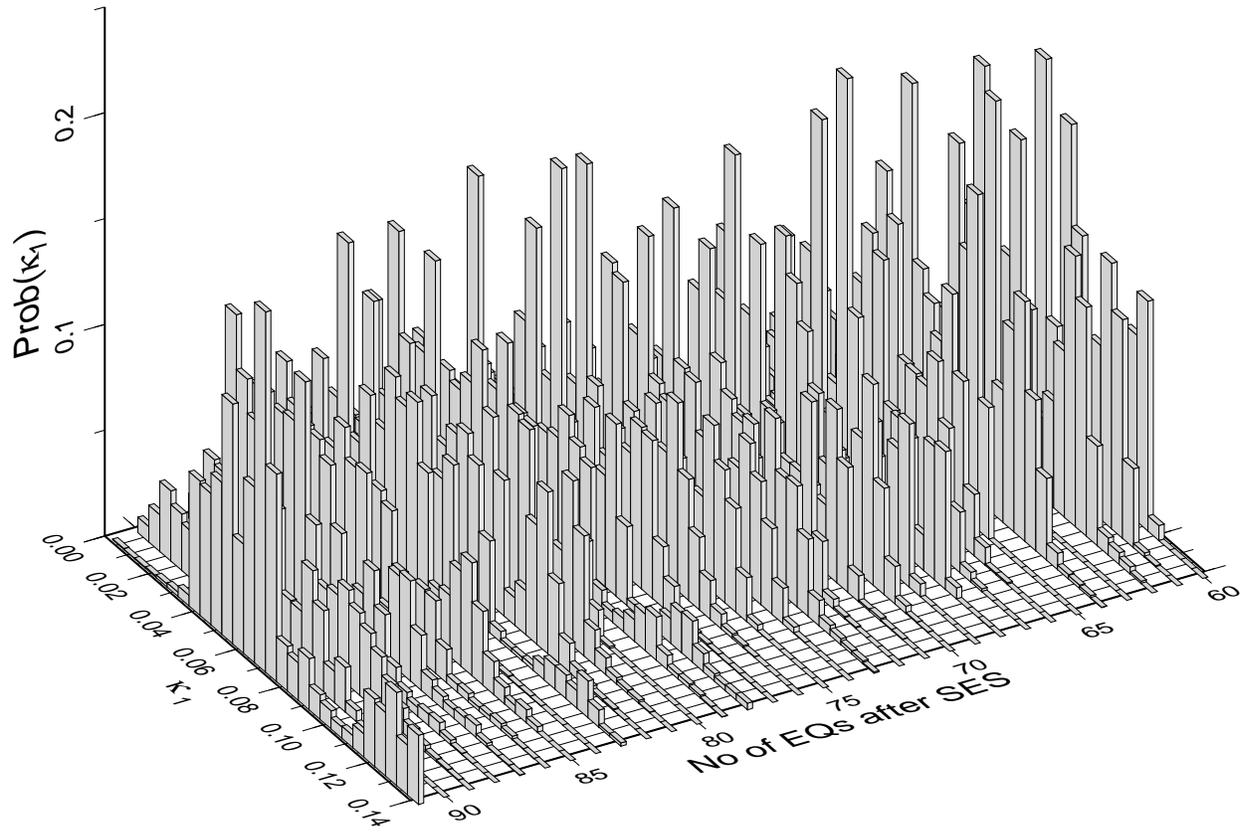}
\caption{The probability Prob($\kappa_1$) vs $\kappa_1$ upon considering
the seismicity (M$_{thres}$=1.9) after January 22, 2011  until 17:10 UT on March 7,
2011 within the area $N_{37.8}^{38.8}E_{22.5}^{24.1}$.} \label{fig9}
\end{figure*}

\begin{figure*}
\includegraphics[scale=0.8,angle=-90]{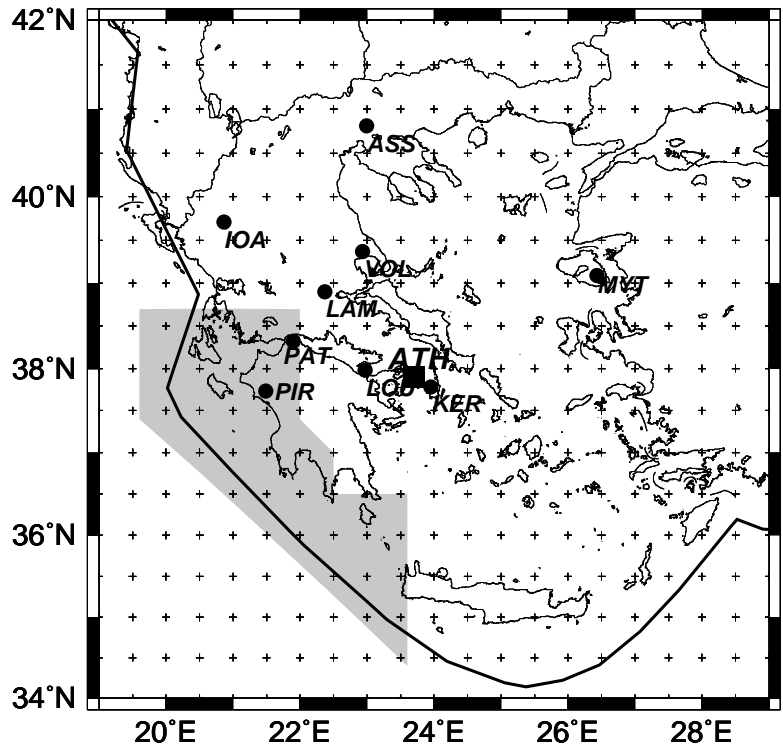}
\caption{The area within which the seismicity was considered in
natural time analysis after the initiation of the SES activity at
PIR lasting from 23:00 UT on May 25 to around 02:00 on May 26,
2011.} \label{fig10}
\end{figure*}

\begin{figure*}
\includegraphics[scale=0.8]{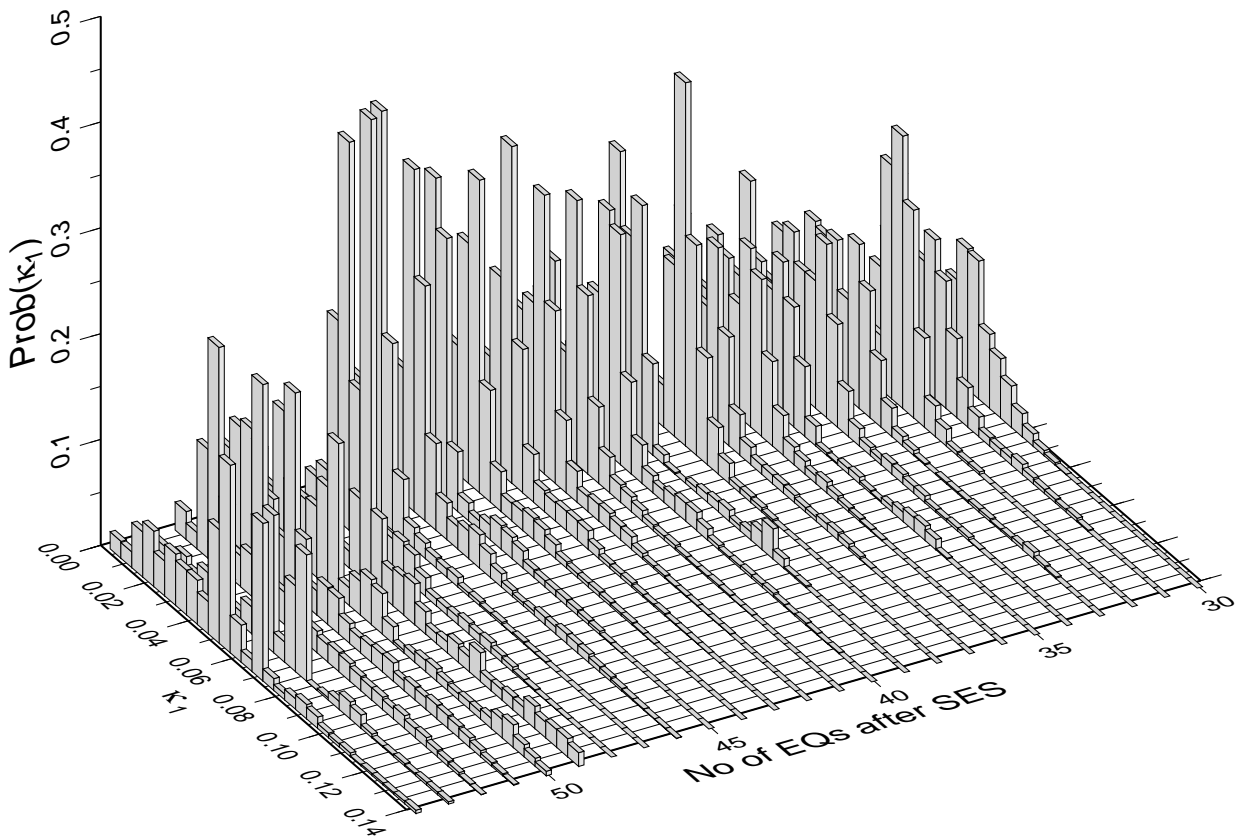}
\caption{The probability Prob($\kappa_1$) vs $\kappa_1$ upon
considering the seismicity (M$_{thres}$=2.7) after the initiation
of the SES activity at PIR (lasting from 23:00 UT on May 25 to
around 02:00 on May 26, 2011) until 07:11 UT on June 24, 2011
within the area depicted in Fig. \ref{fig10}. A maximum of
Prob($\kappa_1$) at $\kappa_1 \approx 0.070$ is also observed for
M$_{thres}$=2.8 upon the occurrence of the ML=4.2 event at 06:53
UT on June 24, 2011.} \label{fig11}
\end{figure*}


\begin{thebibliography}{85}
\expandafter\ifx\csname
natexlab\endcsname\relax\def\natexlab#1{#1}\fi
\expandafter\ifx\csname bibnamefont\endcsname\relax
  \def\bibnamefont#1{#1}\fi
\expandafter\ifx\csname bibfnamefont\endcsname\relax
  \def\bibfnamefont#1{#1}\fi
\expandafter\ifx\csname citenamefont\endcsname\relax
  \def\citenamefont#1{#1}\fi
\expandafter\ifx\csname url\endcsname\relax
  \def\url#1{\texttt{#1}}\fi
\expandafter\ifx\csname
urlprefix\endcsname\relax\def\urlprefix{URL }\fi
\providecommand{\bibinfo}[2]{#2}
\providecommand{\eprint}[2][]{\url{#2}}

\bibitem[{\citenamefont{Bassingthwaighte
  et~al.}(1994)\citenamefont{Bassingthwaighte, Liebovitch, and West}}]{BAS94}
\bibinfo{author}{\bibfnamefont{J.~B.} \bibnamefont{Bassingthwaighte}},
  \bibinfo{author}{\bibfnamefont{L.}~\bibnamefont{Liebovitch}},
  \bibnamefont{and} \bibinfo{author}{\bibfnamefont{B.~J.} \bibnamefont{West}},
  \emph{\bibinfo{title}{Fractal Physiology}} (\bibinfo{publisher}{Oxford
  University Press}, \bibinfo{address}{Oxford UK}, \bibinfo{year}{1994}).

\bibitem[{\citenamefont{M.~Malik and Camm}(1995)}]{MAL95}
\bibinfo{author}{\bibfnamefont{M.}~\bibnamefont{M.~Malik}} \bibnamefont{and}
  \bibinfo{author}{\bibfnamefont{A.}~\bibnamefont{Camm}},
  \emph{\bibinfo{title}{Heart Rate Variability}} (\bibinfo{publisher}{Futura},
  \bibinfo{address}{Armonk NY}, \bibinfo{year}{1995}).

\bibitem[{\citenamefont{Stanley}(1995)}]{STA95}
\bibinfo{author}{\bibfnamefont{H.~E.} \bibnamefont{Stanley}},
  \bibinfo{journal}{Nature} \textbf{\bibinfo{volume}{378}},
  \bibinfo{pages}{554} (\bibinfo{year}{1995}).

\bibitem[{\citenamefont{Hurst}(1951)}]{HUR51}
\bibinfo{author}{\bibfnamefont{H.~E.} \bibnamefont{Hurst}},
  \bibinfo{journal}{Trans. Am. Soc. Civ. Eng.} \textbf{\bibinfo{volume}{116}},
  \bibinfo{pages}{770} (\bibinfo{year}{1951}).

\bibitem[{\citenamefont{Mandelbrot and Wallis}(1969)}]{MAN69}
\bibinfo{author}{\bibfnamefont{B.~B.} \bibnamefont{Mandelbrot}}
  \bibnamefont{and} \bibinfo{author}{\bibfnamefont{J.~R.}
  \bibnamefont{Wallis}}, \bibinfo{journal}{Water Resources Research}
  \textbf{\bibinfo{volume}{5}}, \bibinfo{pages}{321} (\bibinfo{year}{1969}).

\bibitem[{\citenamefont{Stratonovich}(1981)}]{STR81}
\bibinfo{author}{\bibfnamefont{R.~L.} \bibnamefont{Stratonovich}},
  \emph{\bibinfo{title}{Topics in the Theory of Random Noise, vol.I}}
  (\bibinfo{publisher}{Gordon and Breach}, \bibinfo{address}{New York},
  \bibinfo{year}{1981}).

\bibitem[{\citenamefont{Peng et~al.}(1994)\citenamefont{Peng, Buldyrev, Havlin,
  Simons, Stanley, and Goldberger}}]{PEN94}
\bibinfo{author}{\bibfnamefont{C.-K.} \bibnamefont{Peng}},
  \bibinfo{author}{\bibfnamefont{S.~V.} \bibnamefont{Buldyrev}},
  \bibinfo{author}{\bibfnamefont{S.}~\bibnamefont{Havlin}},
  \bibinfo{author}{\bibfnamefont{M.}~\bibnamefont{Simons}},
  \bibinfo{author}{\bibfnamefont{H.~E.} \bibnamefont{Stanley}},
  \bibnamefont{and} \bibinfo{author}{\bibfnamefont{A.~L.}
  \bibnamefont{Goldberger}}, \bibinfo{journal}{Phys. Rev. E}
  \textbf{\bibinfo{volume}{49}}, \bibinfo{pages}{1685} (\bibinfo{year}{1994}).

\bibitem[{\citenamefont{Taqqu et~al.}(1995)\citenamefont{Taqqu, Teverovsky, and
  Willinger}}]{TAQ95}
\bibinfo{author}{\bibfnamefont{M.~S.} \bibnamefont{Taqqu}},
  \bibinfo{author}{\bibfnamefont{V.}~\bibnamefont{Teverovsky}},
  \bibnamefont{and}
  \bibinfo{author}{\bibfnamefont{W.}~\bibnamefont{Willinger}},
  \bibinfo{journal}{Fractals} \textbf{\bibinfo{volume}{3}},
  \bibinfo{pages}{785} (\bibinfo{year}{1995}).

\bibitem[{\citenamefont{Peng et~al.}(1993)\citenamefont{Peng, Buldyrev,
  Goldberger, Havlin, Simons, and Stanley}}]{PEN93}
\bibinfo{author}{\bibfnamefont{C.-K.} \bibnamefont{Peng}},
  \bibinfo{author}{\bibfnamefont{S.~V.} \bibnamefont{Buldyrev}},
  \bibinfo{author}{\bibfnamefont{A.~L.} \bibnamefont{Goldberger}},
  \bibinfo{author}{\bibfnamefont{S.}~\bibnamefont{Havlin}},
  \bibinfo{author}{\bibfnamefont{M.}~\bibnamefont{Simons}}, \bibnamefont{and}
  \bibinfo{author}{\bibfnamefont{H.~E.} \bibnamefont{Stanley}},
  \bibinfo{journal}{Phys. Rev. E} \textbf{\bibinfo{volume}{47}},
  \bibinfo{pages}{3730} (\bibinfo{year}{1993}).

\bibitem[{\citenamefont{Mantegna et~al.}(1994)\citenamefont{Mantegna, Buldyrev,
  Goldberger, Havlin, Peng, Simons, and Stanley}}]{MAN94}
\bibinfo{author}{\bibfnamefont{R.~N.} \bibnamefont{Mantegna}},
  \bibinfo{author}{\bibfnamefont{S.~V.} \bibnamefont{Buldyrev}},
  \bibinfo{author}{\bibfnamefont{A.~L.} \bibnamefont{Goldberger}},
  \bibinfo{author}{\bibfnamefont{S.}~\bibnamefont{Havlin}},
  \bibinfo{author}{\bibfnamefont{C.~K.} \bibnamefont{Peng}},
  \bibinfo{author}{\bibfnamefont{M.}~\bibnamefont{Simons}}, \bibnamefont{and}
  \bibinfo{author}{\bibfnamefont{H.~E.} \bibnamefont{Stanley}},
  \bibinfo{journal}{Phys. Rev. Lett.} \textbf{\bibinfo{volume}{73}},
  \bibinfo{pages}{3169} (\bibinfo{year}{1994}).

\bibitem[{\citenamefont{Havlin et~al.}(1995{\natexlab{a}})\citenamefont{Havlin,
  Buldyrev, Goldberger, Mantegna, Ossadnik, Peng, Simon, and Stanley}}]{HAV95A}
\bibinfo{author}{\bibfnamefont{S.}~\bibnamefont{Havlin}},
  \bibinfo{author}{\bibfnamefont{S.~V.} \bibnamefont{Buldyrev}},
  \bibinfo{author}{\bibfnamefont{A.~L.} \bibnamefont{Goldberger}},
  \bibinfo{author}{\bibfnamefont{R.~N.} \bibnamefont{Mantegna}},
  \bibinfo{author}{\bibfnamefont{S.~M.} \bibnamefont{Ossadnik}},
  \bibinfo{author}{\bibfnamefont{C.~K.} \bibnamefont{Peng}},
  \bibinfo{author}{\bibfnamefont{M.}~\bibnamefont{Simon}}, \bibnamefont{and}
  \bibinfo{author}{\bibfnamefont{H.~E.} \bibnamefont{Stanley}},
  \bibinfo{journal}{Chaos Soliton Fract.} \textbf{\bibinfo{volume}{6}},
  \bibinfo{pages}{171} (\bibinfo{year}{1995}{\natexlab{a}}).

\bibitem[{\citenamefont{Peng et~al.}(1995{\natexlab{a}})\citenamefont{Peng,
  Buldyrev, Goldberger, Havlin, Mantegna, Simon, and Stanley}}]{PEN95}
\bibinfo{author}{\bibfnamefont{C.~K.} \bibnamefont{Peng}},
  \bibinfo{author}{\bibfnamefont{S.~V.} \bibnamefont{Buldyrev}},
  \bibinfo{author}{\bibfnamefont{A.~L.} \bibnamefont{Goldberger}},
  \bibinfo{author}{\bibfnamefont{S.}~\bibnamefont{Havlin}},
  \bibinfo{author}{\bibfnamefont{R.~N.} \bibnamefont{Mantegna}},
  \bibinfo{author}{\bibfnamefont{M.}~\bibnamefont{Simon}}, \bibnamefont{and}
  \bibinfo{author}{\bibfnamefont{H.~E.} \bibnamefont{Stanley}},
  \bibinfo{journal}{Physica A} \textbf{\bibinfo{volume}{221}},
  \bibinfo{pages}{180} (\bibinfo{year}{1995}{\natexlab{a}}).

\bibitem[{\citenamefont{Havlin et~al.}(1995{\natexlab{b}})\citenamefont{Havlin,
  Buldyrev, Goldberger, Mantegna, Peng, Simon, and Stanley}}]{HAV95B}
\bibinfo{author}{\bibfnamefont{S.}~\bibnamefont{Havlin}},
  \bibinfo{author}{\bibfnamefont{S.~V.} \bibnamefont{Buldyrev}},
  \bibinfo{author}{\bibfnamefont{A.~L.} \bibnamefont{Goldberger}},
  \bibinfo{author}{\bibfnamefont{R.~N.} \bibnamefont{Mantegna}},
  \bibinfo{author}{\bibfnamefont{C.~K.} \bibnamefont{Peng}},
  \bibinfo{author}{\bibfnamefont{M.}~\bibnamefont{Simon}}, \bibnamefont{and}
  \bibinfo{author}{\bibfnamefont{H.~E.} \bibnamefont{Stanley}},
  \bibinfo{journal}{Fractals} \textbf{\bibinfo{volume}{3}},
  \bibinfo{pages}{269} (\bibinfo{year}{1995}{\natexlab{b}}).

\bibitem[{\citenamefont{Mantegna et~al.}(1996)\citenamefont{Mantegna, Buldyrev,
  Goldberger, Havlin, Peng, Simons, and Stanley}}]{MAN96}
\bibinfo{author}{\bibfnamefont{R.~N.} \bibnamefont{Mantegna}},
  \bibinfo{author}{\bibfnamefont{S.~V.} \bibnamefont{Buldyrev}},
  \bibinfo{author}{\bibfnamefont{A.~L.} \bibnamefont{Goldberger}},
  \bibinfo{author}{\bibfnamefont{S.}~\bibnamefont{Havlin}},
  \bibinfo{author}{\bibfnamefont{C.-K.} \bibnamefont{Peng}},
  \bibinfo{author}{\bibfnamefont{M.}~\bibnamefont{Simons}}, \bibnamefont{and}
  \bibinfo{author}{\bibfnamefont{H.~E.} \bibnamefont{Stanley}},
  \bibinfo{journal}{Phys. Rev. Lett.} \textbf{\bibinfo{volume}{76}},
  \bibinfo{pages}{1979} (\bibinfo{year}{1996}).

\bibitem[{\citenamefont{Buldyrev et~al.}(1998)\citenamefont{Buldyrev,
  Dokholyan, Goldberger, Havlin, Peng, Stanley, and Viswanathan}}]{BUL98}
\bibinfo{author}{\bibfnamefont{S.~V.} \bibnamefont{Buldyrev}},
  \bibinfo{author}{\bibfnamefont{N.~V.} \bibnamefont{Dokholyan}},
  \bibinfo{author}{\bibfnamefont{A.~L.} \bibnamefont{Goldberger}},
  \bibinfo{author}{\bibfnamefont{S.}~\bibnamefont{Havlin}},
  \bibinfo{author}{\bibfnamefont{C.~K.} \bibnamefont{Peng}},
  \bibinfo{author}{\bibfnamefont{H.~E.} \bibnamefont{Stanley}},
  \bibnamefont{and} \bibinfo{author}{\bibfnamefont{G.~M.}
  \bibnamefont{Viswanathan}}, \bibinfo{journal}{Physica A}
  \textbf{\bibinfo{volume}{249}}, \bibinfo{pages}{430} (\bibinfo{year}{1998}).

\bibitem[{\citenamefont{Stanley et~al.}(1999)\citenamefont{Stanley, Buldyrev,
  Goldberger, Havlin, Peng, and Simon}}]{STA99}
\bibinfo{author}{\bibfnamefont{H.~E.} \bibnamefont{Stanley}},
  \bibinfo{author}{\bibfnamefont{S.~V.} \bibnamefont{Buldyrev}},
  \bibinfo{author}{\bibfnamefont{A.~L.} \bibnamefont{Goldberger}},
  \bibinfo{author}{\bibfnamefont{S.}~\bibnamefont{Havlin}},
  \bibinfo{author}{\bibfnamefont{C.~K.} \bibnamefont{Peng}}, \bibnamefont{and}
  \bibinfo{author}{\bibfnamefont{M.}~\bibnamefont{Simon}},
  \bibinfo{journal}{Physica A} \textbf{\bibinfo{volume}{273}},
  \bibinfo{pages}{1} (\bibinfo{year}{1999}).

\bibitem[{\citenamefont{Peng et~al.}(1995{\natexlab{b}})\citenamefont{Peng,
  Havlin, and Stanley}}]{PEN95B}
\bibinfo{author}{\bibfnamefont{C.~K.} \bibnamefont{Peng}},
  \bibinfo{author}{\bibfnamefont{S.}~\bibnamefont{Havlin}}, \bibnamefont{and}
  \bibinfo{author}{\bibfnamefont{H.~E.} \bibnamefont{Stanley}},
  \bibinfo{journal}{CHAOS} \textbf{\bibinfo{volume}{5}}, \bibinfo{pages}{82}
  (\bibinfo{year}{1995}{\natexlab{b}}).

\bibitem[{\citenamefont{Ho et~al.}(1997)\citenamefont{Ho, Moody, Peng, Mietus,
  Larson, Levy, and Goldberger}}]{HO97}
\bibinfo{author}{\bibfnamefont{K.~K.~L.} \bibnamefont{Ho}},
  \bibinfo{author}{\bibfnamefont{G.~B.} \bibnamefont{Moody}},
  \bibinfo{author}{\bibfnamefont{C.-K.} \bibnamefont{Peng}},
  \bibinfo{author}{\bibfnamefont{J.~E.} \bibnamefont{Mietus}},
  \bibinfo{author}{\bibfnamefont{M.~G.} \bibnamefont{Larson}},
  \bibinfo{author}{\bibfnamefont{D.}~\bibnamefont{Levy}}, \bibnamefont{and}
  \bibinfo{author}{\bibfnamefont{A.~L.} \bibnamefont{Goldberger}},
  \bibinfo{journal}{Circulation} \textbf{\bibinfo{volume}{96}},
  \bibinfo{pages}{842} (\bibinfo{year}{1997}).

\bibitem[{\citenamefont{Ivanov et~al.}(1999)\citenamefont{Ivanov, Bunde,
  Amaral, Havlin, Fritsch-Yelle, Baevsky, Stanley, and Goldberger}}]{IVA99}
\bibinfo{author}{\bibfnamefont{P.~Ch.} \bibnamefont{Ivanov}},
  \bibinfo{author}{\bibfnamefont{A.}~\bibnamefont{Bunde}},
  \bibinfo{author}{\bibfnamefont{L.~A.~N.} \bibnamefont{Amaral}},
  \bibinfo{author}{\bibfnamefont{S.}~\bibnamefont{Havlin}},
  \bibinfo{author}{\bibfnamefont{J.}~\bibnamefont{Fritsch-Yelle}},
  \bibinfo{author}{\bibfnamefont{R.~M.} \bibnamefont{Baevsky}},
  \bibinfo{author}{\bibfnamefont{H.~E.} \bibnamefont{Stanley}},
  \bibnamefont{and} \bibinfo{author}{\bibfnamefont{A.~L.}
  \bibnamefont{Goldberger}}, \bibinfo{journal}{EPL (Europhysics Letters)}
  \textbf{\bibinfo{volume}{48}}, \bibinfo{pages}{594} (\bibinfo{year}{1999}).

\bibitem[{\citenamefont{Ashkenazy et~al.}(2000)\citenamefont{Ashkenazy, Ivanov,
  Havlin, Peng, Yamamoto, Goldberger, and Stanley}}]{ASH00}
\bibinfo{author}{\bibfnamefont{Y.}~\bibnamefont{Ashkenazy}},
  \bibinfo{author}{\bibfnamefont{P.~Ch.} \bibnamefont{Ivanov}},
  \bibinfo{author}{\bibfnamefont{S.}~\bibnamefont{Havlin}},
  \bibinfo{author}{\bibfnamefont{C.~K.} \bibnamefont{Peng}},
  \bibinfo{author}{\bibfnamefont{Y.}~\bibnamefont{Yamamoto}},
  \bibinfo{author}{\bibfnamefont{A.~L.} \bibnamefont{Goldberger}},
  \bibnamefont{and} \bibinfo{author}{\bibfnamefont{H.~E.}
  \bibnamefont{Stanley}}, \bibinfo{journal}{Comput. Cardiol.}
  \textbf{\bibinfo{volume}{27}}, \bibinfo{pages}{139} (\bibinfo{year}{2000}).

\bibitem[{\citenamefont{Ashkenazy et~al.}(2001)\citenamefont{Ashkenazy, Ivanov,
  Havlin, Peng, Goldberger, and Stanley}}]{ASH01}
\bibinfo{author}{\bibfnamefont{Y.}~\bibnamefont{Ashkenazy}},
  \bibinfo{author}{\bibfnamefont{P.~Ch.} \bibnamefont{Ivanov}},
  \bibinfo{author}{\bibfnamefont{S.}~\bibnamefont{Havlin}},
  \bibinfo{author}{\bibfnamefont{C.-K.} \bibnamefont{Peng}},
  \bibinfo{author}{\bibfnamefont{A.~L.} \bibnamefont{Goldberger}},
  \bibnamefont{and} \bibinfo{author}{\bibfnamefont{H.~E.}
  \bibnamefont{Stanley}}, \bibinfo{journal}{Phys. Rev. Lett.}
  \textbf{\bibinfo{volume}{86}}, \bibinfo{pages}{1900} (\bibinfo{year}{2001}).

\bibitem[{\citenamefont{Ivanov et~al.}(2001)\citenamefont{Ivanov, Amaral,
  Goldberger, Halvin, Rosenblum, Stanley, and Struzik}}]{IVA01}
\bibinfo{author}{\bibfnamefont{P.~Ch.} \bibnamefont{Ivanov}},
  \bibinfo{author}{\bibfnamefont{L.~A.~N.} \bibnamefont{Amaral}},
  \bibinfo{author}{\bibfnamefont{A.~L.} \bibnamefont{Goldberger}},
  \bibinfo{author}{\bibfnamefont{S.}~\bibnamefont{Halvin}},
  \bibinfo{author}{\bibfnamefont{M.~G.} \bibnamefont{Rosenblum}},
  \bibinfo{author}{\bibfnamefont{H.~E.} \bibnamefont{Stanley}},
  \bibnamefont{and} \bibinfo{author}{\bibfnamefont{Z.~R.}
  \bibnamefont{Struzik}}, \bibinfo{journal}{CHAOS}
  \textbf{\bibinfo{volume}{11}}, \bibinfo{pages}{641} (\bibinfo{year}{2001}).

\bibitem[{\citenamefont{Kantelhardt
  et~al.}(2002{\natexlab{a}})\citenamefont{Kantelhardt, Ashkenazy, Ivanov,
  Bunde, Havlin, Penzel, Peter, and Stanley}}]{KAN02}
\bibinfo{author}{\bibfnamefont{J.~W.} \bibnamefont{Kantelhardt}},
  \bibinfo{author}{\bibfnamefont{Y.}~\bibnamefont{Ashkenazy}},
  \bibinfo{author}{\bibfnamefont{P.~Ch.} \bibnamefont{Ivanov}},
  \bibinfo{author}{\bibfnamefont{A.}~\bibnamefont{Bunde}},
  \bibinfo{author}{\bibfnamefont{S.}~\bibnamefont{Havlin}},
  \bibinfo{author}{\bibfnamefont{T.}~\bibnamefont{Penzel}},
  \bibinfo{author}{\bibfnamefont{J.-H.} \bibnamefont{Peter}}, \bibnamefont{and}
  \bibinfo{author}{\bibfnamefont{H.~E.} \bibnamefont{Stanley}},
  \bibinfo{journal}{Phys. Rev. E} \textbf{\bibinfo{volume}{65}},
  \bibinfo{pages}{051908} (\bibinfo{year}{2002}{\natexlab{a}}).

\bibitem[{\citenamefont{Karasik et~al.}(2002)\citenamefont{Karasik, Sapir,
  Ashkenazy, Ivanov, Dvir, Lavie, and Havlin}}]{KAR02}
\bibinfo{author}{\bibfnamefont{R.}~\bibnamefont{Karasik}},
  \bibinfo{author}{\bibfnamefont{N.}~\bibnamefont{Sapir}},
  \bibinfo{author}{\bibfnamefont{Y.}~\bibnamefont{Ashkenazy}},
  \bibinfo{author}{\bibfnamefont{P.~Ch.} \bibnamefont{Ivanov}},
  \bibinfo{author}{\bibfnamefont{I.}~\bibnamefont{Dvir}},
  \bibinfo{author}{\bibfnamefont{P.}~\bibnamefont{Lavie}}, \bibnamefont{and}
  \bibinfo{author}{\bibfnamefont{S.}~\bibnamefont{Havlin}},
  \bibinfo{journal}{Phys. Rev. E} \textbf{\bibinfo{volume}{66}},
  \bibinfo{pages}{062902} (\bibinfo{year}{2002}).

\bibitem[{\citenamefont{Ivanov et~al.}(2004)\citenamefont{Ivanov, Chen, Hu, and
  Stanley}}]{IVA04}
\bibinfo{author}{\bibfnamefont{P.~Ch.} \bibnamefont{Ivanov}},
  \bibinfo{author}{\bibfnamefont{Z.}~\bibnamefont{Chen}},
  \bibinfo{author}{\bibfnamefont{K.}~\bibnamefont{Hu}}, \bibnamefont{and}
  \bibinfo{author}{\bibfnamefont{H.~E.} \bibnamefont{Stanley}},
  \bibinfo{journal}{Physica A} \textbf{\bibinfo{volume}{344}},
  \bibinfo{pages}{685} (\bibinfo{year}{2004}).

\bibitem[{\citenamefont{Schmitt and Ivanov}(2007)}]{SCH07}
\bibinfo{author}{\bibfnamefont{D.~T.} \bibnamefont{Schmitt}} \bibnamefont{and}
  \bibinfo{author}{\bibfnamefont{P.~Ch.} \bibnamefont{Ivanov}},
  \bibinfo{journal}{Am. J. Physiol.-Regul. Integr. Comp. Physiol.}
  \textbf{\bibinfo{volume}{293}}, \bibinfo{pages}{R1923}
  (\bibinfo{year}{2007}).

\bibitem[{\citenamefont{Schmitt et~al.}(2009)\citenamefont{Schmitt, Stein, and
  Ivanov}}]{SCH09}
\bibinfo{author}{\bibfnamefont{D.~T.} \bibnamefont{Schmitt}},
  \bibinfo{author}{\bibfnamefont{P.~K.} \bibnamefont{Stein}}, \bibnamefont{and}
  \bibinfo{author}{\bibfnamefont{P.~Ch.} \bibnamefont{Ivanov}},
  \bibinfo{journal}{IEEE Trans. Biomed. Eng.} \textbf{\bibinfo{volume}{56}},
  \bibinfo{pages}{1564} (\bibinfo{year}{2009}).

\bibitem[{\citenamefont{Hu et~al.}(2004)\citenamefont{Hu, Ivanov, Chen, Hilton,
  Stanley, and Shea}}]{HU04}
\bibinfo{author}{\bibfnamefont{K.}~\bibnamefont{Hu}},
  \bibinfo{author}{\bibfnamefont{P.~Ch.} \bibnamefont{Ivanov}},
  \bibinfo{author}{\bibfnamefont{Z.}~\bibnamefont{Chen}},
  \bibinfo{author}{\bibfnamefont{M.~F.} \bibnamefont{Hilton}},
  \bibinfo{author}{\bibfnamefont{H.~E.} \bibnamefont{Stanley}},
  \bibnamefont{and} \bibinfo{author}{\bibfnamefont{S.~A.} \bibnamefont{Shea}},
  \bibinfo{journal}{Physica A} \textbf{\bibinfo{volume}{337}},
  \bibinfo{pages}{307} (\bibinfo{year}{2004}).

\bibitem[{\citenamefont{Ivanov et~al.}(2007)\citenamefont{Ivanov, Hu, Hilton,
  Shea, and Stanley}}]{IVA07A}
\bibinfo{author}{\bibfnamefont{P.~Ch.} \bibnamefont{Ivanov}},
  \bibinfo{author}{\bibfnamefont{K.}~\bibnamefont{Hu}},
  \bibinfo{author}{\bibfnamefont{M.~F.} \bibnamefont{Hilton}},
  \bibinfo{author}{\bibfnamefont{S.~A.} \bibnamefont{Shea}}, \bibnamefont{and}
  \bibinfo{author}{\bibfnamefont{H.~E.} \bibnamefont{Stanley}},
  \bibinfo{journal}{Proc. Natl. Acad. Sci. USA} \textbf{\bibinfo{volume}{104}},
  \bibinfo{pages}{20702} (\bibinfo{year}{2007}).

\bibitem[{\citenamefont{Ivanov}(2007)}]{IVA07B}
\bibinfo{author}{\bibfnamefont{P.~Ch.} \bibnamefont{Ivanov}},
  \bibinfo{journal}{IEEE Eng. Med. Biol.} \textbf{\bibinfo{volume}{26}},
  \bibinfo{pages}{33} (\bibinfo{year}{2007}).

\bibitem[{\citenamefont{Hu et~al.}(2007)\citenamefont{Hu, Scheer, Ivanov,
  Buijs, and Shea}}]{HU07}
\bibinfo{author}{\bibfnamefont{K.}~\bibnamefont{Hu}},
  \bibinfo{author}{\bibfnamefont{F.~A.~J.~L.} \bibnamefont{Scheer}},
  \bibinfo{author}{\bibfnamefont{P.~Ch.} \bibnamefont{Ivanov}},
  \bibinfo{author}{\bibfnamefont{R.~M.} \bibnamefont{Buijs}}, \bibnamefont{and}
  \bibinfo{author}{\bibfnamefont{S.~A.} \bibnamefont{Shea}},
  \bibinfo{journal}{Neuroscience} \textbf{\bibinfo{volume}{149}},
  \bibinfo{pages}{508} (\bibinfo{year}{2007}).

\bibitem[{\citenamefont{Ivanova and Ausloos}(1999)}]{IVAN99}
\bibinfo{author}{\bibfnamefont{K.}~\bibnamefont{Ivanova}} \bibnamefont{and}
  \bibinfo{author}{\bibfnamefont{M.}~\bibnamefont{Ausloos}},
  \bibinfo{journal}{Physica A} \textbf{\bibinfo{volume}{274}},
  \bibinfo{pages}{349} (\bibinfo{year}{1999}).

\bibitem[{\citenamefont{Koscielny-Bunde
  et~al.}(1998)\citenamefont{Koscielny-Bunde, Bunde, Havlin, Roman, Goldreich,
  and Schellnhuber}}]{BUN98}
\bibinfo{author}{\bibfnamefont{E.}~\bibnamefont{Koscielny-Bunde}},
  \bibinfo{author}{\bibfnamefont{A.}~\bibnamefont{Bunde}},
  \bibinfo{author}{\bibfnamefont{S.}~\bibnamefont{Havlin}},
  \bibinfo{author}{\bibfnamefont{H.~E.} \bibnamefont{Roman}},
  \bibinfo{author}{\bibfnamefont{Y.}~\bibnamefont{Goldreich}},
  \bibnamefont{and} \bibinfo{author}{\bibfnamefont{H.-J.}
  \bibnamefont{Schellnhuber}}, \bibinfo{journal}{Phys. Rev. Lett.}
  \textbf{\bibinfo{volume}{81}}, \bibinfo{pages}{729} (\bibinfo{year}{1998}).

\bibitem[{\citenamefont{Talkner and Weber}(2000)}]{TAL00}
\bibinfo{author}{\bibfnamefont{P.}~\bibnamefont{Talkner}} \bibnamefont{and}
  \bibinfo{author}{\bibfnamefont{R.~O.} \bibnamefont{Weber}},
  \bibinfo{journal}{Phys. Rev. E} \textbf{\bibinfo{volume}{62}},
  \bibinfo{pages}{150} (\bibinfo{year}{2000}).

\bibitem[{\citenamefont{Bunde et~al.}(2001)\citenamefont{Bunde, Havlin,
  Koscielny-Bunde, and Schellnhuber}}]{BUN01}
\bibinfo{author}{\bibfnamefont{A.}~\bibnamefont{Bunde}},
  \bibinfo{author}{\bibfnamefont{S.}~\bibnamefont{Havlin}},
  \bibinfo{author}{\bibfnamefont{E.}~\bibnamefont{Koscielny-Bunde}},
  \bibnamefont{and} \bibinfo{author}{\bibfnamefont{H.~J.}
  \bibnamefont{Schellnhuber}}, \bibinfo{journal}{Physica A}
  \textbf{\bibinfo{volume}{302}}, \bibinfo{pages}{255} (\bibinfo{year}{2001}).

\bibitem[{\citenamefont{Monetti et~al.}(2003)\citenamefont{Monetti, Havlin, and
  Bunde}}]{MON03}
\bibinfo{author}{\bibfnamefont{R.~A.} \bibnamefont{Monetti}},
  \bibinfo{author}{\bibfnamefont{S.}~\bibnamefont{Havlin}}, \bibnamefont{and}
  \bibinfo{author}{\bibfnamefont{A.}~\bibnamefont{Bunde}},
  \bibinfo{journal}{Physica A} \textbf{\bibinfo{volume}{320}},
  \bibinfo{pages}{581} (\bibinfo{year}{2003}).

\bibitem[{\citenamefont{Bunde et~al.}(2005)\citenamefont{Bunde, Eichner,
  Kantelhardt, and Havlin}}]{BUN05}
\bibinfo{author}{\bibfnamefont{A.}~\bibnamefont{Bunde}},
  \bibinfo{author}{\bibfnamefont{J.~F.} \bibnamefont{Eichner}},
  \bibinfo{author}{\bibfnamefont{J.~W.} \bibnamefont{Kantelhardt}},
  \bibnamefont{and} \bibinfo{author}{\bibfnamefont{S.}~\bibnamefont{Havlin}},
  \bibinfo{journal}{Phys. Rev. Lett.} \textbf{\bibinfo{volume}{94}},
  \bibinfo{pages}{048701} (\bibinfo{year}{2005}).

\bibitem[{\citenamefont{Liu et~al.}(1997)\citenamefont{Liu, Cizeau, Meyer,
  Peng, and Stanley}}]{LIU97}
\bibinfo{author}{\bibfnamefont{Y.~H.} \bibnamefont{Liu}},
  \bibinfo{author}{\bibfnamefont{P.}~\bibnamefont{Cizeau}},
  \bibinfo{author}{\bibfnamefont{M.}~\bibnamefont{Meyer}},
  \bibinfo{author}{\bibfnamefont{C.~K.} \bibnamefont{Peng}}, \bibnamefont{and}
  \bibinfo{author}{\bibfnamefont{H.~E.} \bibnamefont{Stanley}},
  \bibinfo{journal}{Physica A} \textbf{\bibinfo{volume}{245}},
  \bibinfo{pages}{437} (\bibinfo{year}{1997}).

\bibitem[{\citenamefont{Vandewalle and Ausloos}(1997)}]{VAN97}
\bibinfo{author}{\bibfnamefont{N.}~\bibnamefont{Vandewalle}} \bibnamefont{and}
  \bibinfo{author}{\bibfnamefont{M.}~\bibnamefont{Ausloos}},
  \bibinfo{journal}{Physica A} \textbf{\bibinfo{volume}{246}},
  \bibinfo{pages}{454} (\bibinfo{year}{1997}).

\bibitem[{\citenamefont{Vandewalle and Ausloos}(1998)}]{VAN98}
\bibinfo{author}{\bibfnamefont{N.}~\bibnamefont{Vandewalle}} \bibnamefont{and}
  \bibinfo{author}{\bibfnamefont{M.}~\bibnamefont{Ausloos}},
  \bibinfo{journal}{Phys. Rev. E} \textbf{\bibinfo{volume}{58}},
  \bibinfo{pages}{6832} (\bibinfo{year}{1998}).

\bibitem[{\citenamefont{Ausloos et~al.}(1999)\citenamefont{Ausloos, Vandewalle,
  Boveroux, Minguet, and Ivanova}}]{AUS99}
\bibinfo{author}{\bibfnamefont{M.}~\bibnamefont{Ausloos}},
  \bibinfo{author}{\bibfnamefont{N.}~\bibnamefont{Vandewalle}},
  \bibinfo{author}{\bibfnamefont{P.}~\bibnamefont{Boveroux}},
  \bibinfo{author}{\bibfnamefont{A.}~\bibnamefont{Minguet}}, \bibnamefont{and}
  \bibinfo{author}{\bibfnamefont{K.}~\bibnamefont{Ivanova}},
  \bibinfo{journal}{Physica A} \textbf{\bibinfo{volume}{274}},
  \bibinfo{pages}{229} (\bibinfo{year}{1999}).

\bibitem[{\citenamefont{Vandewalle et~al.}(1999)\citenamefont{Vandewalle,
  Ausloos, and Boveroux}}]{VAN99}
\bibinfo{author}{\bibfnamefont{N.}~\bibnamefont{Vandewalle}},
  \bibinfo{author}{\bibfnamefont{M.}~\bibnamefont{Ausloos}}, \bibnamefont{and}
  \bibinfo{author}{\bibfnamefont{P.}~\bibnamefont{Boveroux}},
  \bibinfo{journal}{Physica A} \textbf{\bibinfo{volume}{269}},
  \bibinfo{pages}{170} (\bibinfo{year}{1999}).

\bibitem[{\citenamefont{Ausloos}(2000)}]{AUS00}
\bibinfo{author}{\bibfnamefont{M.}~\bibnamefont{Ausloos}},
  \bibinfo{journal}{Physica A} \textbf{\bibinfo{volume}{285}},
  \bibinfo{pages}{48} (\bibinfo{year}{2000}).

\bibitem[{\citenamefont{Ausloos and Ivanova}(2001)}]{AUS01}
\bibinfo{author}{\bibfnamefont{M.}~\bibnamefont{Ausloos}} \bibnamefont{and}
  \bibinfo{author}{\bibfnamefont{K.}~\bibnamefont{Ivanova}},
  \bibinfo{journal}{Phys. Rev. E} \textbf{\bibinfo{volume}{63}},
  \bibinfo{pages}{047201} (\bibinfo{year}{2001}).

\bibitem[{\citenamefont{Varotsos
  et~al.}(2003{\natexlab{a}})\citenamefont{Varotsos, Sarlis, and
  Skordas}}]{NAT03A}
\bibinfo{author}{\bibfnamefont{P.~A.} \bibnamefont{Varotsos}},
  \bibinfo{author}{\bibfnamefont{N.~V.} \bibnamefont{Sarlis}},
  \bibnamefont{and} \bibinfo{author}{\bibfnamefont{E.~S.}
  \bibnamefont{Skordas}}, \bibinfo{journal}{Phys. Rev. E}
  \textbf{\bibinfo{volume}{67}}, \bibinfo{pages}{021109}
  (\bibinfo{year}{2003}{\natexlab{a}}).

\bibitem[{\citenamefont{Varotsos
  et~al.}(2003{\natexlab{b}})\citenamefont{Varotsos, Sarlis, and
  Skordas}}]{NAT03B}
\bibinfo{author}{\bibfnamefont{P.~A.} \bibnamefont{Varotsos}},
  \bibinfo{author}{\bibfnamefont{N.~V.} \bibnamefont{Sarlis}},
  \bibnamefont{and} \bibinfo{author}{\bibfnamefont{E.~S.}
  \bibnamefont{Skordas}}, \bibinfo{journal}{Phys. Rev. E}
  \textbf{\bibinfo{volume}{68}}, \bibinfo{pages}{031106}
  (\bibinfo{year}{2003}{\natexlab{b}}).

\bibitem[{\citenamefont{Varotsos et~al.}(2009)\citenamefont{Varotsos, Sarlis,
  and Skordas}}]{NAT09}
\bibinfo{author}{\bibfnamefont{P.~A.} \bibnamefont{Varotsos}},
  \bibinfo{author}{\bibfnamefont{N.~V.} \bibnamefont{Sarlis}},
  \bibnamefont{and} \bibinfo{author}{\bibfnamefont{E.~S.}
  \bibnamefont{Skordas}}, \bibinfo{journal}{CHAOS}
  \textbf{\bibinfo{volume}{19}}, \bibinfo{eid}{023114} (\bibinfo{year}{2009}).

\bibitem[{\citenamefont{Varotsos and Alexopoulos}(1984{\natexlab{a}})}]{VAR84A}
\bibinfo{author}{\bibfnamefont{P.}~\bibnamefont{Varotsos}} \bibnamefont{and}
  \bibinfo{author}{\bibfnamefont{K.}~\bibnamefont{Alexopoulos}},
  \bibinfo{journal}{Tectonophysics} \textbf{\bibinfo{volume}{110}},
  \bibinfo{pages}{73} (\bibinfo{year}{1984}{\natexlab{a}}).

\bibitem[{\citenamefont{Varotsos and Alexopoulos}(1984{\natexlab{b}})}]{VAR84B}
\bibinfo{author}{\bibfnamefont{P.}~\bibnamefont{Varotsos}} \bibnamefont{and}
  \bibinfo{author}{\bibfnamefont{K.}~\bibnamefont{Alexopoulos}},
  \bibinfo{journal}{Tectonophysics} \textbf{\bibinfo{volume}{110}},
  \bibinfo{pages}{99} (\bibinfo{year}{1984}{\natexlab{b}}).

\bibitem[{\citenamefont{Varotsos et~al.}(1986)\citenamefont{Varotsos,
  Alexopoulos, Nomicos, and Lazaridou}}]{VAR86}
\bibinfo{author}{\bibfnamefont{P.}~\bibnamefont{Varotsos}},
  \bibinfo{author}{\bibfnamefont{K.}~\bibnamefont{Alexopoulos}},
  \bibinfo{author}{\bibfnamefont{K.}~\bibnamefont{Nomicos}}, \bibnamefont{and}
  \bibinfo{author}{\bibfnamefont{M.}~\bibnamefont{Lazaridou}},
  \bibinfo{journal}{Nature (London)} \textbf{\bibinfo{volume}{322}},
  \bibinfo{pages}{120} (\bibinfo{year}{1986}).

\bibitem[{\citenamefont{Varotsos et~al.}(1988)\citenamefont{Varotsos,
  Alexopoulos, Nomicos, and Lazaridou}}]{VAR88}
\bibinfo{author}{\bibfnamefont{P.}~\bibnamefont{Varotsos}},
  \bibinfo{author}{\bibfnamefont{K.}~\bibnamefont{Alexopoulos}},
  \bibinfo{author}{\bibfnamefont{K.}~\bibnamefont{Nomicos}}, \bibnamefont{and}
  \bibinfo{author}{\bibfnamefont{M.}~\bibnamefont{Lazaridou}},
  \bibinfo{journal}{Tectonophysics} \textbf{\bibinfo{volume}{152}},
  \bibinfo{pages}{193} (\bibinfo{year}{1988}).

\bibitem[{\citenamefont{Varotsos and Lazaridou}(1991)}]{VAR91}
\bibinfo{author}{\bibfnamefont{P.}~\bibnamefont{Varotsos}} \bibnamefont{and}
  \bibinfo{author}{\bibfnamefont{M.}~\bibnamefont{Lazaridou}},
  \bibinfo{journal}{Tectonophysics} \textbf{\bibinfo{volume}{188}},
  \bibinfo{pages}{321} (\bibinfo{year}{1991}).

\bibitem[{\citenamefont{Varotsos et~al.}(1993)\citenamefont{Varotsos,
  Alexopoulos, and Lazaridou}}]{VAR93A}
\bibinfo{author}{\bibfnamefont{P.}~\bibnamefont{Varotsos}},
  \bibinfo{author}{\bibfnamefont{K.}~\bibnamefont{Alexopoulos}},
  \bibnamefont{and}
  \bibinfo{author}{\bibfnamefont{M.}~\bibnamefont{Lazaridou}},
  \bibinfo{journal}{Tectonophysics} \textbf{\bibinfo{volume}{224}},
  \bibinfo{pages}{1} (\bibinfo{year}{1993}).

\bibitem[{\citenamefont{Varotsos
  et~al.}(1996{\natexlab{a}})\citenamefont{Varotsos, Eftaxias, Lazaridou,
  Antonopoulos, Makris, and Poliyiannakis}}]{VAR96A}
\bibinfo{author}{\bibfnamefont{P.}~\bibnamefont{Varotsos}},
  \bibinfo{author}{\bibfnamefont{K.}~\bibnamefont{Eftaxias}},
  \bibinfo{author}{\bibfnamefont{M.}~\bibnamefont{Lazaridou}},
  \bibinfo{author}{\bibfnamefont{G.}~\bibnamefont{Antonopoulos}},
  \bibinfo{author}{\bibfnamefont{J.}~\bibnamefont{Makris}}, \bibnamefont{and}
  \bibinfo{author}{\bibfnamefont{J.}~\bibnamefont{Poliyiannakis}},
  \bibinfo{journal}{Geophys. Res. Lett.} \textbf{\bibinfo{volume}{23}},
  \bibinfo{pages}{1449} (\bibinfo{year}{1996}{\natexlab{a}}).

\bibitem[{\citenamefont{Varotsos
  et~al.}(2003{\natexlab{c}})\citenamefont{Varotsos, Sarlis, and
  Skordas}}]{PRL03}
\bibinfo{author}{\bibfnamefont{P.~A.} \bibnamefont{Varotsos}},
  \bibinfo{author}{\bibfnamefont{N.~V.} \bibnamefont{Sarlis}},
  \bibnamefont{and} \bibinfo{author}{\bibfnamefont{E.~S.}
  \bibnamefont{Skordas}}, \bibinfo{journal}{Phys. Rev. Lett.}
  \textbf{\bibinfo{volume}{91}}, \bibinfo{pages}{148501}
  (\bibinfo{year}{2003}{\natexlab{c}}).

\bibitem[{\citenamefont{Sarlis and Varotsos}(2002)}]{SAR02}
\bibinfo{author}{\bibfnamefont{N.}~\bibnamefont{Sarlis}} \bibnamefont{and}
  \bibinfo{author}{\bibfnamefont{P.}~\bibnamefont{Varotsos}},
  \bibinfo{journal}{J. Geodyn.} \textbf{\bibinfo{volume}{33}},
  \bibinfo{pages}{463} (\bibinfo{year}{2002}).

\bibitem[{\citenamefont{Varotsos
  et~al.}(2001{\natexlab{a}})\citenamefont{Varotsos, Sarlis, and
  Skordas}}]{VAR01A}
\bibinfo{author}{\bibfnamefont{P.}~\bibnamefont{Varotsos}},
  \bibinfo{author}{\bibfnamefont{N.}~\bibnamefont{Sarlis}}, \bibnamefont{and}
  \bibinfo{author}{\bibfnamefont{E.}~\bibnamefont{Skordas}},
  \bibinfo{journal}{Proc. Jpn. Acad., Ser. B: Phys. Biol. Sci.}
  \textbf{\bibinfo{volume}{77}}, \bibinfo{pages}{87}
  (\bibinfo{year}{2001}{\natexlab{a}}).

\bibitem[{\citenamefont{Varotsos
  et~al.}(2001{\natexlab{b}})\citenamefont{Varotsos, Sarlis, and
  Skordas}}]{VAR01B}
\bibinfo{author}{\bibfnamefont{P.}~\bibnamefont{Varotsos}},
  \bibinfo{author}{\bibfnamefont{N.}~\bibnamefont{Sarlis}}, \bibnamefont{and}
  \bibinfo{author}{\bibfnamefont{E.}~\bibnamefont{Skordas}},
  \bibinfo{journal}{Proc. Jpn. Acad., Ser. B: Phys. Biol. Sci.}
  \textbf{\bibinfo{volume}{77}}, \bibinfo{pages}{93}
  (\bibinfo{year}{2001}{\natexlab{b}}).

\bibitem[{\citenamefont{Weber and Talkner}(2001)}]{WEB01}
\bibinfo{author}{\bibfnamefont{R.~O.} \bibnamefont{Weber}} \bibnamefont{and}
  \bibinfo{author}{\bibfnamefont{P.}~\bibnamefont{Talkner}},
  \bibinfo{journal}{J. Geophys. Res.-Atmos} \textbf{\bibinfo{volume}{106}},
  \bibinfo{pages}{20131} (\bibinfo{year}{2001}).

\bibitem[{\citenamefont{Kantelhardt
  et~al.}(2002{\natexlab{b}})\citenamefont{Kantelhardt, Zschiegner,
  Koscienly-Bunde, Bunde, Havlin, and Stanley}}]{KAN02B}
\bibinfo{author}{\bibfnamefont{J.}~\bibnamefont{Kantelhardt}},
  \bibinfo{author}{\bibfnamefont{S.~A.} \bibnamefont{Zschiegner}},
  \bibinfo{author}{\bibfnamefont{E.}~\bibnamefont{Koscienly-Bunde}},
  \bibinfo{author}{\bibfnamefont{A.}~\bibnamefont{Bunde}},
  \bibinfo{author}{\bibfnamefont{S.}~\bibnamefont{Havlin}}, \bibnamefont{and}
  \bibinfo{author}{\bibfnamefont{H.~E.} \bibnamefont{Stanley}},
  \bibinfo{journal}{Physica A} \textbf{\bibinfo{volume}{316}},
  \bibinfo{pages}{87} (\bibinfo{year}{2002}{\natexlab{b}}).

\bibitem[{\citenamefont{Muzy et~al.}(1994)\citenamefont{Muzy, Bacry, and
  Arneodo}}]{MUZ94}
\bibinfo{author}{\bibfnamefont{J.~F.} \bibnamefont{Muzy}},
  \bibinfo{author}{\bibfnamefont{E.}~\bibnamefont{Bacry}}, \bibnamefont{and}
  \bibinfo{author}{\bibfnamefont{A.}~\bibnamefont{Arneodo}},
  \bibinfo{journal}{Int. J. Bifurcation Chaos} \textbf{\bibinfo{volume}{4}},
  \bibinfo{pages}{245} (\bibinfo{year}{1994}).

\bibitem[{\citenamefont{Varotsos and Alexopoulos}(1986)}]{VARBOOK}
\bibinfo{author}{\bibfnamefont{P.}~\bibnamefont{Varotsos}} \bibnamefont{and}
  \bibinfo{author}{\bibfnamefont{K.}~\bibnamefont{Alexopoulos}},
  \emph{\bibinfo{title}{Thermodynamics of Point Defects and their Relation with
  Bulk Properties}} (\bibinfo{publisher}{North Holland},
  \bibinfo{address}{Amsterdam}, \bibinfo{year}{1986}).

\bibitem[{\citenamefont{Varotsos}(2005)}]{NEWBOOK}
\bibinfo{author}{\bibfnamefont{P.}~\bibnamefont{Varotsos}},
  \emph{\bibinfo{title}{The Physics of Seismic Electric Signals}}
  (\bibinfo{publisher}{TERRAPUB}, \bibinfo{address}{Tokyo},
  \bibinfo{year}{2005}).

\bibitem[{\citenamefont{Varotsos}(1976)}]{VAR76}
\bibinfo{author}{\bibfnamefont{P.}~\bibnamefont{Varotsos}},
  \bibinfo{journal}{Phys. Rev. B} \textbf{\bibinfo{volume}{13}},
  \bibinfo{pages}{938} (\bibinfo{year}{1976}).

\bibitem[{\citenamefont{Varotsos and Alexopoulos}(1978)}]{VAR78A}
\bibinfo{author}{\bibfnamefont{P.}~\bibnamefont{Varotsos}} \bibnamefont{and}
  \bibinfo{author}{\bibfnamefont{K.}~\bibnamefont{Alexopoulos}},
  \bibinfo{journal}{J. Phys. Chem. Solids} \textbf{\bibinfo{volume}{39}},
  \bibinfo{pages}{759} (\bibinfo{year}{1978}).

\bibitem[{\citenamefont{Varotsos and Alexopoulos}(1979)}]{VAR79A}
\bibinfo{author}{\bibfnamefont{P.}~\bibnamefont{Varotsos}} \bibnamefont{and}
  \bibinfo{author}{\bibfnamefont{K.}~\bibnamefont{Alexopoulos}},
  \bibinfo{journal}{J. Physics C: Solid State} \textbf{\bibinfo{volume}{12}},
  \bibinfo{pages}{L761} (\bibinfo{year}{1979}).

\bibitem[{\citenamefont{Varotsos and Alexopoulos}(1984{\natexlab{c}})}]{VAR84C}
\bibinfo{author}{\bibfnamefont{P.}~\bibnamefont{Varotsos}} \bibnamefont{and}
  \bibinfo{author}{\bibfnamefont{K.}~\bibnamefont{Alexopoulos}},
  \bibinfo{journal}{Phys. Rev. B} \textbf{\bibinfo{volume}{30}},
  \bibinfo{pages}{7305} (\bibinfo{year}{1984}{\natexlab{c}}).

\bibitem[{\citenamefont{Varotsos}(2007)}]{VAR07}
\bibinfo{author}{\bibfnamefont{P.}~\bibnamefont{Varotsos}},
  \bibinfo{journal}{J. Appl. Phys.} \textbf{\bibinfo{volume}{101}},
  \bibinfo{eid}{123503} (\bibinfo{year}{2007}).

\bibitem[{\citenamefont{Varotsos et~al.}(1978)\citenamefont{Varotsos, Ludwig,
  and Alexopoulos}}]{VAR78B}
\bibinfo{author}{\bibfnamefont{P.}~\bibnamefont{Varotsos}},
  \bibinfo{author}{\bibfnamefont{W.}~\bibnamefont{Ludwig}}, \bibnamefont{and}
  \bibinfo{author}{\bibfnamefont{K.}~\bibnamefont{Alexopoulos}},
  \bibinfo{journal}{Phys. Rev. B} \textbf{\bibinfo{volume}{18}},
  \bibinfo{pages}{2683} (\bibinfo{year}{1978}).

\bibitem[{\citenamefont{Ma et~al.}(2010)\citenamefont{Ma, Bartsch,
  Bernaola-Galv\'an, Yoneyama, and Ivanov}}]{MA10}
\bibinfo{author}{\bibfnamefont{Q.~D.~Y.} \bibnamefont{Ma}},
  \bibinfo{author}{\bibfnamefont{R.~P.} \bibnamefont{Bartsch}},
  \bibinfo{author}{\bibfnamefont{P.}~\bibnamefont{Bernaola-Galv\'an}},
  \bibinfo{author}{\bibfnamefont{M.}~\bibnamefont{Yoneyama}}, \bibnamefont{and}
  \bibinfo{author}{\bibfnamefont{P.~Ch.} \bibnamefont{Ivanov}}
  (\bibinfo{year}{2010}), \eprint{arXiv.org:physics.data-an/1001.3641v2}.

\bibitem[{\citenamefont{Orihara et~al.}(2009)\citenamefont{Orihara, Kamogawa,
  Nagao, and Uyeda}}]{ORI09}
\bibinfo{author}{\bibfnamefont{Y.}~\bibnamefont{Orihara}},
  \bibinfo{author}{\bibfnamefont{M.}~\bibnamefont{Kamogawa}},
  \bibinfo{author}{\bibfnamefont{T.}~\bibnamefont{Nagao}}, \bibnamefont{and}
  \bibinfo{author}{\bibfnamefont{S.}~\bibnamefont{Uyeda}},
  \bibinfo{journal}{Proc. Jpn. Acad., Ser. B: Phys. Biol. Sci.}
  \textbf{\bibinfo{volume}{85}}, \bibinfo{pages}{435} (\bibinfo{year}{2009}).

\bibitem[{\citenamefont{Uyeda et~al.}(2009)\citenamefont{Uyeda, Kamogawa, and
  Tanaka}}]{UYE09}
\bibinfo{author}{\bibfnamefont{S.}~\bibnamefont{Uyeda}},
  \bibinfo{author}{\bibfnamefont{M.}~\bibnamefont{Kamogawa}}, \bibnamefont{and}
  \bibinfo{author}{\bibfnamefont{H.}~\bibnamefont{Tanaka}},
  \bibinfo{journal}{J. Geophys. Res.} \textbf{\bibinfo{volume}{114}},
  \bibinfo{pages}{B02310, doi:10.1029/2007JB005332} (\bibinfo{year}{2009}).

\bibitem[{\citenamefont{Varotsos et~al.}(2002)\citenamefont{Varotsos, Sarlis,
  and Skordas}}]{NAT02}
\bibinfo{author}{\bibfnamefont{P.~A.} \bibnamefont{Varotsos}},
  \bibinfo{author}{\bibfnamefont{N.~V.} \bibnamefont{Sarlis}},
  \bibnamefont{and} \bibinfo{author}{\bibfnamefont{E.~S.}
  \bibnamefont{Skordas}}, \bibinfo{journal}{Phys. Rev. E}
  \textbf{\bibinfo{volume}{66}}, \bibinfo{pages}{011902}
  (\bibinfo{year}{2002}).

\bibitem[{\citenamefont{Varotsos et~al.}(2004)\citenamefont{Varotsos, Sarlis,
  Skordas, and Lazaridou}}]{NAT04}
\bibinfo{author}{\bibfnamefont{P.~A.} \bibnamefont{Varotsos}},
  \bibinfo{author}{\bibfnamefont{N.~V.} \bibnamefont{Sarlis}},
  \bibinfo{author}{\bibfnamefont{E.~S.} \bibnamefont{Skordas}},
  \bibnamefont{and} \bibinfo{author}{\bibfnamefont{M.~S.}
  \bibnamefont{Lazaridou}}, \bibinfo{journal}{Phys. Rev. E}
  \textbf{\bibinfo{volume}{70}}, \bibinfo{pages}{011106}
  (\bibinfo{year}{2004}).

\bibitem[{\citenamefont{Varotsos
  et~al.}(2005{\natexlab{a}})\citenamefont{Varotsos, Sarlis, Tanaka, and
  Skordas}}]{NAT05B}
\bibinfo{author}{\bibfnamefont{P.~A.} \bibnamefont{Varotsos}},
  \bibinfo{author}{\bibfnamefont{N.~V.} \bibnamefont{Sarlis}},
  \bibinfo{author}{\bibfnamefont{H.~K.} \bibnamefont{Tanaka}},
  \bibnamefont{and} \bibinfo{author}{\bibfnamefont{E.~S.}
  \bibnamefont{Skordas}}, \bibinfo{journal}{Phys. Rev. E}
  \textbf{\bibinfo{volume}{71}}, \bibinfo{pages}{032102}
  (\bibinfo{year}{2005}{\natexlab{a}}).

\bibitem[{\citenamefont{Varotsos
  et~al.}(2005{\natexlab{b}})\citenamefont{Varotsos, Sarlis, Skordas, and
  Lazaridou}}]{NAT05A}
\bibinfo{author}{\bibfnamefont{P.~A.} \bibnamefont{Varotsos}},
  \bibinfo{author}{\bibfnamefont{N.~V.} \bibnamefont{Sarlis}},
  \bibinfo{author}{\bibfnamefont{E.~S.} \bibnamefont{Skordas}},
  \bibnamefont{and} \bibinfo{author}{\bibfnamefont{M.~S.}
  \bibnamefont{Lazaridou}}, \bibinfo{journal}{Phys. Rev. E}
  \textbf{\bibinfo{volume}{71}}, \bibinfo{pages}{011110}
  (\bibinfo{year}{2005}{\natexlab{b}}).

\bibitem[{\citenamefont{Varotsos
  et~al.}(2006{\natexlab{a}})\citenamefont{Varotsos, Sarlis, Skordas, Tanaka,
  and Lazaridou}}]{NAT06A}
\bibinfo{author}{\bibfnamefont{P.~A.} \bibnamefont{Varotsos}},
  \bibinfo{author}{\bibfnamefont{N.~V.} \bibnamefont{Sarlis}},
  \bibinfo{author}{\bibfnamefont{E.~S.} \bibnamefont{Skordas}},
  \bibinfo{author}{\bibfnamefont{H.~K.} \bibnamefont{Tanaka}},
  \bibnamefont{and} \bibinfo{author}{\bibfnamefont{M.~S.}
  \bibnamefont{Lazaridou}}, \bibinfo{journal}{Phys. Rev. E}
  \textbf{\bibinfo{volume}{73}}, \bibinfo{pages}{031114}
  (\bibinfo{year}{2006}{\natexlab{a}}).

\bibitem[{\citenamefont{Varotsos
  et~al.}(2006{\natexlab{b}})\citenamefont{Varotsos, Sarlis, Skordas, Tanaka,
  and Lazaridou}}]{NAT06B}
\bibinfo{author}{\bibfnamefont{P.~A.} \bibnamefont{Varotsos}},
  \bibinfo{author}{\bibfnamefont{N.~V.} \bibnamefont{Sarlis}},
  \bibinfo{author}{\bibfnamefont{E.~S.} \bibnamefont{Skordas}},
  \bibinfo{author}{\bibfnamefont{H.~K.} \bibnamefont{Tanaka}},
  \bibnamefont{and} \bibinfo{author}{\bibfnamefont{M.~S.}
  \bibnamefont{Lazaridou}}, \bibinfo{journal}{Phys. Rev. E}
  \textbf{\bibinfo{volume}{74}}, \bibinfo{pages}{021123}
  (\bibinfo{year}{2006}{\natexlab{b}}).

\bibitem[{\citenamefont{Lesche}(1982)}]{LES82}
\bibinfo{author}{\bibfnamefont{B.}~\bibnamefont{Lesche}}, \bibinfo{journal}{J.
  Stat. Phys.} \textbf{\bibinfo{volume}{27}}, \bibinfo{pages}{419}
  (\bibinfo{year}{1982}).

\bibitem[{\citenamefont{Lesche}(2004)}]{LES04}
\bibinfo{author}{\bibfnamefont{B.}~\bibnamefont{Lesche}},
  \bibinfo{journal}{Phys. Rev. E} \textbf{\bibinfo{volume}{70}},
  \bibinfo{pages}{017102} (\bibinfo{year}{2004}).

\bibitem[{\citenamefont{Varotsos et~al.}(2010)\citenamefont{Varotsos, Sarlis,
  and Skordas}}]{ARXIV10}
\bibinfo{author}{\bibfnamefont{P.~A.} \bibnamefont{Varotsos}},
  \bibinfo{author}{\bibfnamefont{N.~V.} \bibnamefont{Sarlis}},
  \bibnamefont{and} \bibinfo{author}{\bibfnamefont{E.~S.}
  \bibnamefont{Skordas}} (\bibinfo{year}{2010}),
  \eprint{arXiv.org:cond-mat.stat-mech/0904.2465v10}.

\bibitem[{kei()}]{keimeno}
\bibinfo{note}{The SES activity of Fig. \ref{fig1}(b) originated the natural
  time analysis of the seismicity in the area N(38.0-39.0) E(21.5-23.7) after
  December 27, 2009, described in the previous reference. This type of analysis
  suggests that the critical point has been approached, which reveals that an
  impending mainshock, if any, is imminent (See also Ref.[86]).}

\bibitem[{\citenamefont{Varotsos
  et~al.}(1996{\natexlab{b}})\citenamefont{Varotsos, Lazaridou, Eftaxias,
  Antonopoulos, Makris, and Kopanas}}]{VAR96B}
\bibinfo{author}{\bibfnamefont{P.}~\bibnamefont{Varotsos}},
  \bibinfo{author}{\bibfnamefont{M.}~\bibnamefont{Lazaridou}},
  \bibinfo{author}{\bibfnamefont{K.}~\bibnamefont{Eftaxias}},
  \bibinfo{author}{\bibfnamefont{G.}~\bibnamefont{Antonopoulos}},
  \bibinfo{author}{\bibfnamefont{J.}~\bibnamefont{Makris}}, \bibnamefont{and}
  \bibinfo{author}{\bibfnamefont{J.}~\bibnamefont{Kopanas}}, in
  \emph{\bibinfo{booktitle}{The Critical Review of VAN: Earthquake Prediction
  from Seismic Electric Signals}}, edited by
  \bibinfo{editor}{\bibfnamefont{Sir ~J.} \bibnamefont{Lighthill}}
  (\bibinfo{publisher}{World Scientific}, \bibinfo{address}{Singapore},
  \bibinfo{year}{1996}{\natexlab{b}}), pp. \bibinfo{pages}{29--76}.

\bibitem[{\citenamefont{Lighthill}(1996)}]{LIG96}
\bibinfo{author}{\bibfnamefont{J.}~\bibnamefont{Lighthill}}, in
  \emph{\bibinfo{booktitle}{The Critical Review of VAN: Earthquake Prediction
  from Seismic Electric Signals}}, edited by
  \bibinfo{editor}{\bibfnamefont{Sir ~J.} \bibnamefont{Lighthill}}
  (\bibinfo{publisher}{World Scientific}, \bibinfo{address}{Singapore},
  \bibinfo{year}{1996}), pp. \bibinfo{pages}{373--376}.

\bibitem[{\citenamefont{Sarlis et~al.}(2008)\citenamefont{Sarlis, Skordas,
  Lazaridou, and Varotsos}}]{SAR08}
\bibinfo{author}{\bibfnamefont{N.~V.} \bibnamefont{Sarlis}},
  \bibinfo{author}{\bibfnamefont{E.~S.} \bibnamefont{Skordas}},
  \bibinfo{author}{\bibfnamefont{M.~S.} \bibnamefont{Lazaridou}},
  \bibnamefont{and} \bibinfo{author}{\bibfnamefont{P.~A.}
  \bibnamefont{Varotsos}}, \bibinfo{journal}{Proc. Jpn. Acad., Ser. B: Phys.
  Biol. Sci.} \textbf{\bibinfo{volume}{84}}, \bibinfo{pages}{331}
  (\bibinfo{year}{2008}).

\bibitem[{kei()}]{keimeno2}
\bibinfo{note}{{\em Note added on April 16, 2010:} Actually, as proposed in Ref.[82], three
days later, i.e., on March 9, 2010, a Ms(ATH)=5.6 earthquake occurred at 38.87N 23.65E
inside the expected area. In addition, upon repeating the natural time analysis
of seismicity$^{85}$ in the area N(37.65-39.0) E(22.2-24.1) (KER) after February 25, 2010,
we again find an approach to {\em criticality} in the same fashion as described in Ref.[82].
The condition for criticality was found to hold only for magnitude thresholds $M_{thres}=2.8$
and $M_{thres}=2.9$ and hence in order to assure magnitude threshold invariance further study
is in progress.}

\bibitem[{kei()}]{keimeno3}
\bibinfo{note}{{\em Note added on May 29, 2010:} The aforementioned study in the area  N(37.65-39.0) E(22.2-24.1) finally revealed magnitude threshold invariance of the {\em criticality} condition upon the occurrence of the ML=3.0 event at 14:32 UT on May 27, 2010, with an epicenter at 38.5N 23.4E. This can be seen when plotting the probability Prob($\kappa_1$) versus $\kappa_1$ for $M_{thres}=2.8$, 2.9 and 3.0 in accordance with the procedure described in Ref.[85] identifying the occurrence time of an impending mainshock.}


\bibitem[{kei()}]{keimeno4}
\bibinfo{note}{{\em Note added on August 23, 2010:} The previous
Note was followed on July 16, 2010 by Ms(ATH)=5.6 earthquake at
39.3N 24.0E, i.e., very close to the northeastern edge of the
predicted area. Two other SES activities were recorded at Patras
station (PAT) at 09:07 UT on August 9, 2010 and at 06:24 UT on
August 10, 2010 depicted in Figs. 5(a) and 5(b), respectively.
They resulted in an analysis, in natural time, of the seismicity
in the area N(37.6-39.0) E(20.4-23.2) similar to that in
Ref.[82].}

\bibitem[{kei()}]{keimeno5}
\bibinfo{note}{{\em Note added on October 1, 2010:} This is just a continuation of the analysis started, as mentioned in the previous Note, in the area N(37.6-39.0) E(20.4-23.2) which is now extended to N(37.5-38.8) E(19.8-23.3). At 03:53 UT on September 2, 2010 a Ms(ATH)=4.8 earthquake actually occurred with an epicenter at 38.22N 23.17E, i.e., inside the area studied. In addition, upon the occurrence of a M$_L=$3.1 event at 15:42 UT on September 29,
2010  with an epicenter at 37.56N 20.00E, when plotting the probability Prob($\kappa_1$) versus $\kappa_1$, for M$_{thres}$=3.1, we find a sharp maximum at $\kappa_1\approx 0.070$ depicted in Fig.\ref{fig6}, which probably indicates that the critical point has been approached. To ensure that this is a true {\em criticality} condition, the extent to which a magnitude threshold invariance holds for this result is currently investigated.}

\bibitem[{kei()}]{keimeno6}
\bibinfo{note}{{\em Note added on October 28, 2010:} Actually, almost one week after the previous Note, i.e., at 19:04 UT on October 9, 2010, a Ms(ATH)=4.8  earthquake occurred with an epicenter at 38.15N 22.72E lying inside the area studied. The continuation of the study related to the SES activities recorded at PAT on August 9 \& 10, 2010 (Fig.\ref{fig5}) in the same area N(37.5-38.8) E(19.8-23.3), reveals the following:
Upon the occurrence of a ML=4.1 earthquake at 10:32 UT on October 23, 2010, with an epicenter at 38.73N 21.99E, a maximum in the plot of Prob($\kappa_1$) versus $\kappa_1$ for M$_{thres}$=3.3 is observed at $\kappa_1=0.070$. The same was observed for   M$_{thres}$=3.2 upon the occurrence of a ML=3.2 earthquake at 00:38 UT on October 24, 2010, with an epicenter at 38.71N 21.99E as well as for   M$_{thres}$=3.1 upon the occurrence of a ML=3.6 earthquake at 04:04 UT on October 28, 2010, with an epicenter at 38.36N 22.25E. These findings are  strikingly reminiscent of the following behavior: For higher magnitude threshold, the description of the real situation {\em approaching criticality} becomes less accurate due to ``coarse graining''. The plot for the aforementioned maximum observed on October 28, 2010 is depicted in Fig.\ref{fig7}.}

\bibitem[{kei()}]{keimeno7}
\bibinfo{note}{{\em Note added on January 27, 2011:} At $\approx$ 01:00 UT on January 22, 2011 an electric disturbance
started at KER which lasted for around 9 hours. This does not obey
the $\Delta V / L$ criterion so that to be classified as SES activity. On
the other hand, when analyzing it in natural time we find the
following values $\kappa_1=0.072(6)$, $S=0.075(6)$ and $S_-=0.085(9)$ which conform with
the properties of a true precursory signal. A violation however of
the $\Delta V / L$ criterion may be due to the inhomogeneities in
that area, and in order to resolve the possibility of an approach to criticality, we continued the investigation described in the previous Note as well as in the area $N_{37.8}^{38.8}$ $E_{22.8}^{24.1}$. We find that (for M$_{thres}$=3.2) upon the occurrence of a ML=3.2 event at 09:31 UT on January 24, 2011, a maximum of Prob($\kappa_1$) versus $\kappa_1$ is observed at $\kappa_1=0.070$, see Fig.\ref{fig8}, which possibly indicates the approach of the system to the critical point.}

\bibitem[{kei()}]{keimeno8}
\bibinfo{note}{{\em Note added on March 8, 2011:} Upon the occurrence of the ML=3.5 event at 17:10 UT on March 7, 2011, the Prob($\kappa_1$) versus $\kappa_1$ exhibited a maximum at the value  $\kappa_1=0.070$ for M$_{thres}$=1.1 to 1.9 (see Fig. \ref{fig9} for M$_{thres}$=1.9). If the seismic data for such low magnitude thresholds are complete (which is questionable), the present behavior suggests that the critical point is approached (i.e., the event is expected within a few days to around 1 week or so) in the area $N_{37.8}^{38.8}E_{22.5}^{24.1}$. Since $\Delta V / L$ is of the order of 5mV/km,  Ms(ATH) depending on the epicentral distance may reach 6 {\em if} $r\sim $100km.}


\bibitem[{kei()}]{keimeno9}
\bibinfo{note}{{\em Note added on June 24, 2011:} Since the
seismic data in the previous Note for such low magnitude
thresholds were {\em not} complete, as questioned, the time-window
calculation could not be accurate. An earthquake of magnitude
M$_s$(ATH)=4.4 actually occurred later on April 22, 2011 with an
epicenter at 38.37$^o$N 23.62$^o$E, i.e., around 60-70km from KER.
In addition, a SES activity was recorded at PIR lasting for around
3 hours, i.e., from 23:00 UT on May 25 to around 02:00 on May 26,
2011. Thus, a natural time analysis of the subsequent seismicity
started in the shaded area depicted in Fig.\ref{fig10} in a
similar fashion as in arXiv:0802.3329v4 (publicized on May 29,
2008 before the 6.5 on June 8, 2008) following the same procedure
described in the latter publication. This revealed that, upon the
occurrence of a ML=2.9 earthquake at 07:11 UT on June 24, 2011
with an epicenter at 37.6$^o$N 21.0$^o$E, the Prob($\kappa_1$)
versus $\kappa_1$ exhibited a maximum at $\kappa_1 \approx 0.070$
(as shown in Fig. \ref{fig11}), thus pointing to the approach of
the system to the critical point.}

\bibitem[{kei()}]{keimeno10}
\bibinfo{note}{{\em Note added on January 26, 2012:} In a recent manuscript  submitted for publication on January 23, 2012, the recording of a strong electric disturbance at PIR (see Fig.\ref{fig10}) on January 6, 2012 has been reported. A copy of its recording along with the current investigation on its tentative classification as SES activity can be found elsewhere (see the file {\tt arxiv\_230112.pdf} available from  \url{http://physlab.phys.uoa.gr/org/director.htm}). }

\end{thebibliography}

\end{document}